\documentclass[12pt]{article}
\usepackage{graphicx,rotating,epsfig,amsmath,amssymb,color,multirow,pstricks,pst-plot}
\newlength{\digitwidth} 

\newcommand{\mlab}[1]%
    {\mbox{}\marginpar{\raggedright\hspace{0pt}\footnotesize #1}}

\newcommand{\pt}{$p_{\mathrm{T}}$}

\newcommand{\xf}{$x_{\mathrm F}$}

\newcommand{\dd}{{\mathrm d}}
\newcommand{\ccbar}{${c}\bar{c}$}

\newcommand{\qqbar}{$q\bar{q}$}

\newcommand{\sabs}{$\sigma_{\rm abs}$}

\newcommand{\jpsi}{J/$\psi$}
\newcommand{\psip}{$\psi^\prime$}
\newcommand{\chic}{$\chi_c$}

\newcommand{\sabsjpsi}{$\sigma_{\rm abs}^{\rm J/\psi}$}

\newcommand{\ssnn}{$\sqrt{s_{_{NN}}}$}
\newcommand{\sspsi}{$\sqrt{s_{_{\psi N}}}$}

\newcommand\jhep[3]{{{\it J.\ High\ Energy\ Phys.\ }{\bf #1} (#2) #3}}
\newcommand\epjc[3]{{{\it Eur.\ Phys.\ J. }{\bf C #1} (#2) #3}}

\newcommand\plb[3] {{{\it Phys.\ Lett.\ }{\bf B #1} (#2) #3}}
\newcommand\prc[3] {{{\it Phys.\ Rev.\ }{\bf C #1} (#2) #3}}
\newcommand\prd[3] {{{\it Phys.\ Rev.\ }{\bf D #1} (#2) #3}}
\newcommand\prl[3] {{{\it Phys.\ Rev.\ Lett.\ }{\bf #1} (#2) #3}}
\newcommand\zpc[3] {{{\it Z.\ Physik }{\bf C #1} (#2) #3}}
\newcommand\prep[3]{{{\it Phys.\ Rept.\ }{\bf #1} (#2) #3}}
\newcommand\ijmpa[3]{{{\it Int.\ J.\ Mod.\ Phys.\ }{\bf A #1} (#2) #3}}

\let \oldref=\ref
\renewcommand{\ref}[1]{{\red\oldref{#1}}}

\begin{document}

\begingroup
\thispagestyle{empty}
\baselineskip=14pt
\parskip 0pt plus 5pt

\begin{center}
{\large EUROPEAN LABORATORY FOR PARTICLE PHYSICS}
\end{center}

\bigskip
\begin{flushright}
CERN--PH--EP\,/\,2008--019\\
November 12, 2008
\end{flushright}

\bigskip
\begin{center}
{\Large\bf\boldmath
Energy dependence of \jpsi\ absorption\\[0.3 cm]
in proton-nucleus collisions}

\bigskip\bigskip

Carlos Louren\c{c}o\\
CERN-PH, CH-1211 Geneva 23, {\it carlos.lourenco@cern.ch}\\[0.5cm]

Ramona Vogt\\
LLNL, Livermore, CA 94550, USA, {\it vogt2@llnl.gov}\\
and 
University of California, Davis, CA, USA\\[0.5cm]

Hermine K. W\"ohri\\
LIP, Lisbon, Portugal, {\it hermine.woehri@cern.ch}\\[0.5cm]

\textbf{Abstract}

\end{center}

\begingroup
\leftskip=0.4cm
\rightskip=0.4cm
\parindent=0.pt

  Charmonium states are expected to be considerably suppressed
  in the case of quark-gluon plasma formation in high-energy heavy-ion
  collisions.  However, a robust identification of suppression
  patterns as signatures of a deconfined QCD medium requires a
  detailed understanding of the ``normal nuclear absorption'' already
  present in proton-nucleus collisions, where the charmonium
  production cross sections increase less than linearly with the
  number of target nucleons.
  We analyse the \jpsi\ production cross sections measured in
  proton-nucleus collisions in fixed target experiments, with proton
  beam energies from 200 to 920~GeV, and in d-Au collisions at RHIC,
  at $\sqrt{s_{_{NN}}}=200$~GeV, in the framework of the Glauber
  formalism, using several sets of parton distributions with and
  without nuclear modifications.
  The results reveal a significant dependence of the ``absorption
  cross section'' on the kinematics of the \jpsi\ and on the collision
  energy.  Extrapolating the observed patterns we derive the level of
  absorption expected at $E_{\rm lab} = 158$~GeV, the energy at which
  the heavy-ion data sets were collected at the CERN SPS.

\bigskip

Keywords: \jpsi\ absorption, cold nuclear matter effects


\endgroup

\vfill
\begin{center}
\emph{To be published in JHEP}
\end{center}

\endgroup

\newpage
\thispagestyle{empty}
~
\tableofcontents

\newpage
\pagenumbering{arabic}
\setcounter{page}{1}

\section{Introduction}

According to lattice QCD calculations~\cite{karsch}, when hadronic
matter reaches sufficiently high energy densities it should undergo a
phase transition to a ``plasma'' of deconfined quarks and gluons (the
QGP phase).  Considerable efforts are currently being invested in the
study of high-energy heavy-ion collisions to reveal the existence of
this phase transition and to study the properties of the new phase in
view of improving our understanding of confinement, a crucial feature
of QCD.
The production yields of the quarkonium states should be considerably
suppressed by ``colour screening'' if a QCD medium with deconfined
quarks and gluons is formed in high-energy heavy-ion
collisions~\cite{MatsuiSatz}.  However, already in proton-nucleus
collisions the charmonium production cross sections increase less than
linearly with the number of binary nucleon-nucleon collisions.  This
``normal nuclear absorption'' needs to be well understood so that a
robust baseline reference can be established, with respect to which we
can clearly and unambiguously identify the signals of ``new physics''
specific to the high density QCD medium.

So far, the studies of the \jpsi\ suppression patterns determined by
the NA50 and NA60 experiments at the SPS, from data collected with Pb
and In ion beams of 158~GeV per nucleon, use a baseline reference
established on the basis of proton-nucleus measurements performed at
400 and 450~GeV, assuming that the energy dependencies of the initial
and final state ``normal nuclear effects'', if any, can be neglected.
It is important to verify if this simple assumption is supported or
not by existing experimental evidence, looking at the results reported
by experiments made at different energies, complementing and placing
in a broader context the results to be obtained by NA60 from
proton-nucleus measurements made at 158~GeV.

In this paper we present a detailed study of some ``cold nuclear
matter effects'' affecting charmonium production in proton-nucleus
collisions, as well as their energy dependence.  Section~\ref{sec:history}
gives a brief historical motivation.  Section~\ref{sec:theory}
describes the basic framework of our study: how the charmonium
production cross sections are calculated and how the absorption of the
charmonium states in the nuclear matter is evaluated.  Comparing our
calculations to the midrapidity \jpsi\ production cross sections
measured in proton-nucleus collisions by NA3 (200~GeV), NA50
(400/450~GeV), E866 (800~GeV) and \mbox{HERA-B} (920~GeV), presented
in Section~\ref{sec:data}, we observe a significant energy dependence
of \sabsjpsi\ in this energy range.  Simple parametrisations of this
energy dependence lead to \jpsi\ normal nuclear absorption rates at
the SPS heavy-ion energy, $E_{\rm lab}=158$~GeV, significantly larger
than at higher energies, as presented and discussed in
Section~\ref{sec:analysis}.

\section{Brief historical motivation}
\label{sec:history}

The NA50 experiment at the CERN-SPS made a detailed study of \jpsi\
and \psip\ production in fixed-target proton-nucleus collisions with
incident protons of 400 and 450~GeV, employing six different nuclear
targets (Be, Al, Cu, Ag, W and Pb)~\cite{NA50-400, NA50-450}.  
Comparing the production cross sections measured at 400~GeV, for
instance, to calculations based on the Glauber formalism (neglecting
nuclear modifications of the parton densities), the \jpsi\
``absorption cross section'' was determined to be $\sigma_{\rm
  abs}^{\rm J/\psi} = 4.6 \pm 0.6$~mb~\cite{NA50-400}.
A similar value, $\sigma_{\rm abs}^{\rm J/\psi} = 4.2 \pm 0.5$~mb, is
extracted from a global fit to the 400 and 450~GeV ${\rm
  J}/\psi~/~{\rm DY}$ cross-section \emph{ratios}, where Drell-Yan
dimuons are used as reference.
This value has been used by NA50~\cite{NA50PbPb} and NA60~\cite{NA60}
in the studies of the SPS heavy-ion data, collected at 158~GeV,
assuming that the energy dependencies of the initial and final state
``normal nuclear effects'', if any, can be neglected.

An attempt was made~\cite{Goncalo} to check the reliability of
this assumption, exploring the centrality dependence of the ${\rm
  J}/\psi~/~{\rm DY}$ cross-section ratio measured by the NA38
experiment in S-U collisions at 200~GeV per nucleon.  This study gave
$\sigma_{\rm abs}^{\rm J/\psi} = 7.1 \pm 2.8$~mb, a value larger than
the one derived from the higher energy data.  However, this
result is not satisfactory given the large uncertainty in the value of
$\sigma_{\rm abs}^{\rm J/\psi}$ and the need to assume that there are no 
additional nuclear
effects between the proton-nucleus and the S-U collision systems.  We
know, in particular, that the \psip\ state is considerably more
suppressed in S-U than in p-A collisions~\cite{psiprime}, and we
should not assume that the \jpsi\ state is insensitive to the additional
mechanisms of \psip\ absorption.

Whether or not charmonium absorption depends on the collision energy still
remains an open question.  It is long known that
E866, at FNAL, observed less \jpsi\ absorption at 800~GeV than seen by
NA50 in the same $x_{\rm F}\sim 0$ region.  In terms of the very
simple ``$\alpha$ parametrisation'',
\begin{equation}
  \sigma_{\rm p\,A} = \sigma_{0} \times A^\alpha \quad ,
\label{eq:alpha}
\end{equation}
NA50 reported $\alpha=0.925\pm 0.009$ at 400~GeV~\cite{NA50-400} while
E866 obtained values around 0.95~\cite{E866}.

At the other end of the energy scale, NA3 reported $\alpha \sim 0.94$ 
at $x_{\rm F}\sim 0$~\cite{NA3}.  However, it has meanwhile been
observed~\cite{Ruben} that an absorption pattern generated
using the Glauber framework, with a certain absorption cross section,
leads to significantly different $\alpha$ values when the light target
is Hydrogen (used by NA3) or Beryllium (used by E866).  In other
words, if NA3 had used Be as the light target, as did NA50 and
E866, they would have obtained an $\alpha$ value around 0.92.

A global average of \jpsi\ absorption cross sections was recently
reported~\cite{Arleo}, assuming that measurements collected with
different beam particles and energies should reflect a single \sabs\
value despite the observation that some of the values
were mutually exclusive.

Knowing the crucial importance of the normal nuclear absorption
baseline in the interpretation of the \jpsi\ suppression seen in the
heavy-ion data and given that the charmonium absorption processes may
very well depend on the collision energy~\cite{na60-2004}, the NA60
experiment collected (in 2004) proton-nucleus collisions at 158~GeV,
the energy of the Pb and In beams used by NA50 and NA60, with seven
different nuclear targets (Be, Al, Cu, In, W, Pb and U).  In addition,
the PHENIX experiment at RHIC should soon report measurements of
\jpsi\ absorption in $\sqrt{s_{_{NN}}} = 200$~GeV d-Au collisions with
much better accuracy than that available in Ref.~\cite{PHENIX}.
These two sets of results, obtained at very different energies, should
help determine the existence of an energy dependence and significantly
improve our understanding of the mechanisms causing the observed
nuclear effects in charmonium production.

\section{Basic elements of the calculations}
\label{sec:theory}

This section describes the framework we have used to derive the \jpsi\
absorption cross sections.  The calculations of the charmonium
production cross sections are performed with the colour evaporation
model (CEM) using several sets of parton distribution functions (PDFs).  It should
be emphasised, however, that our results are derived by studying how
the production yields change from light to heavy targets and,
therefore, are essentially insensitive to the specific production
model and set of PDFs used.  We performed calculations with ``free
proton'' PDFs and also with PDFs modified by the nuclear environment,
using several parametrisations of the nuclear modifications.  The
survival probability of the charmonium states traversing the nuclear
matter was evaluated in the framework of the Glauber model.

\subsection{Charmonium production cross sections}
\label{sec:cem}

The charmonium production cross sections used in the studies reported
in this paper were calculated using the colour evaporation model,
described in detail in Ref.~\cite{HPC}.  In the CEM, the production
cross section of each charmonium state, $\tilde{\sigma}_i$, is assumed
to be a constant fraction, $F_i$, of the total ``closed charm''
production cross section, calculated as the integral over the \ccbar\
pair mass, $m$, of the \ccbar\ cross section, from threshold, $2\,m_c$,
to twice the mass of the lightest D
meson, $2 \, m_{\rm D}=3.74$~GeV/$c^2$:
\begin{equation}
\frac{\dd \tilde{\sigma}_i}{\dd x_{\rm F}} = 2F_i
\int_{2m_c}^{2m_{\rm D}} m \, \dd m \, \frac{\dd \sigma^{c \bar c}}{\dd
x_{\rm F} \,\dd m^2} \quad ,
\label{eq:cevap}
\end{equation}
where $x_{\rm F}$ is the \ccbar\ longitudinal momentum fraction in the
centre-of-mass frame of the two colliding hadrons.

At leading order in perturbative QCD, the \ccbar\ hadroproduction
cross section is given by the sum of two partonic contributions, gluon
fusion ($gg$) and quark-antiquark annihilation (\qqbar), convoluted
with the parton densities in the colliding hadrons, $A$ and
$B$~\cite{Bkp2}:
\begin{eqnarray}
\label{eq:cemdef}
\frac{\dd \sigma^{c \bar c}}{\dd x_{\rm F} \,\dd m^2} 
& = & \int_0^1 \dd x_1\, \dd x_2 \, \delta(x_1x_2 s_{_{NN}} - m^2) \,
\delta ( x_{\rm F} - x_1 + x_2 )\nonumber \\
& & \quad\quad\quad\quad\quad \Big\{ f_g^A(x_1,m^2)\, f_g^B(x_2,m^2)\, \sigma_{gg}(m^2) + \\
& & \sum_{q=u,d,s} [f_q^A(x_1,m^2) f_{\bar q}^B(x_2,m^2) +
f_{\bar q}^A(x_1,m^2) f_q^B(x_2,m^2)] \sigma_{q \bar q}(m^2)\Big\} \nonumber \quad ,
\end{eqnarray} 
where $x_1$ and $x_2$ are the momentum fractions carried by the two
partons and \ssnn\ is the centre-of-mass energy of the
nucleon-nucleon collision.  In our calculations we use leading order
parton densities, evaluated at scale $m^2 = x_1\, x_2\, s_{_{NN}}$.

A basic feature of the CEM is that $\dd \sigma^{c \bar c}/\dd x_{\rm F} \, \dd
m^2$ fully determines the energy and momentum dependencies of all the states.
The hadronization of the \ccbar\ pairs into charmonium states is
nonperturbative, involving the emission of one or more soft gluons.
Each state requires a different matrix element, condensed in the fractions
$F_i$, calculated for
each state: J/$\psi$, $\psi'$, $\chi_{c_J}$, etc.  We note that $F_{\rm
  J/\psi}$ includes both direct \jpsi\ production and feed-down \jpsi\
production through $\chi_{c_J}$ radiative decays and \psip\ hadronic
decays.

\subsection{Parton densities in the proton and in the nucleus}
\label{sec:nPDFs}

Deep inelastic scattering (DIS) and Drell-Yan measurements performed
with nuclear targets have shown that the distributions of partons in
nuclei are significantly modified relative to those in free protons.
These nuclear modifications depend on the fraction of the total hadron
momentum carried by the parton, $x$, on the momentum scale, $Q^2$, and
on the mass number of the nucleus, $A$.  While the mechanisms governing these
modifications are not yet well understood, several groups have produced
parametrisations, $S_i(A,x,Q^2)$, that convert
the free-proton distributions for each parton $i$, $f_i^{\rm
  p}(x,Q^2)$, into nuclear ones, $f_i^A(x,Q^2)$, assuming factorisation:
\begin{equation}
f_i^A(x,Q^2) = S_i(A,x,Q^2) \times f_i^{\rm p}(x,Q^2) \quad .
\label{eq:npdfs}
\end{equation}
Naturally, the nuclear modifications should depend on the spatial
location of the nucleon inside the target, with the nucleons at the
surface being less ``shadowed'' than those in the core of the
nucleus~\cite{ekkv}.  This effect can be ignored in analyses of
p-nucleus data samples integrated over collision centrality (as done
in this work).

\begin{figure}[ht]\centering
\resizebox{0.48\textwidth}{!}{%
\includegraphics*{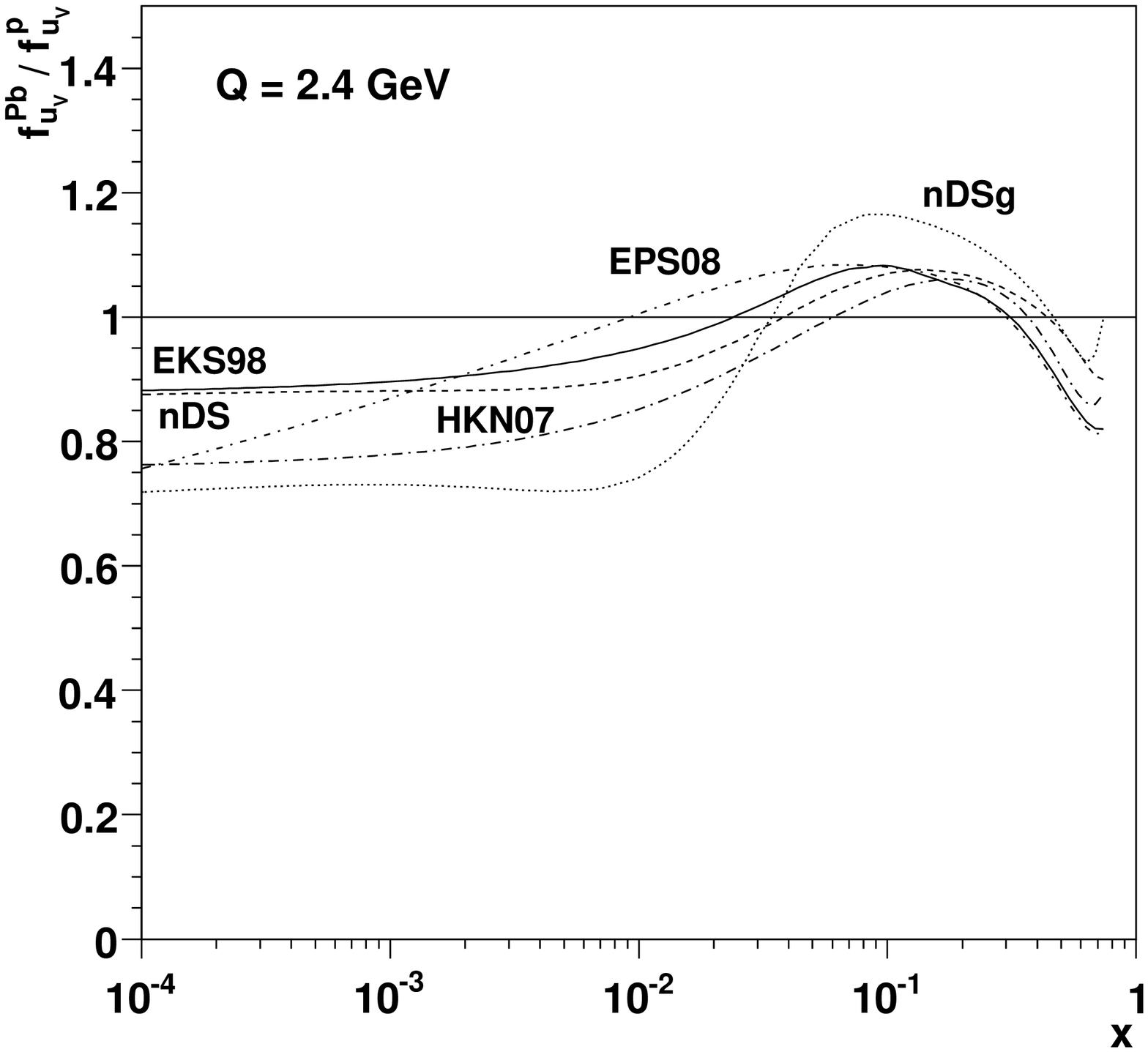}}
\resizebox{0.48\textwidth}{!}{%
\includegraphics*{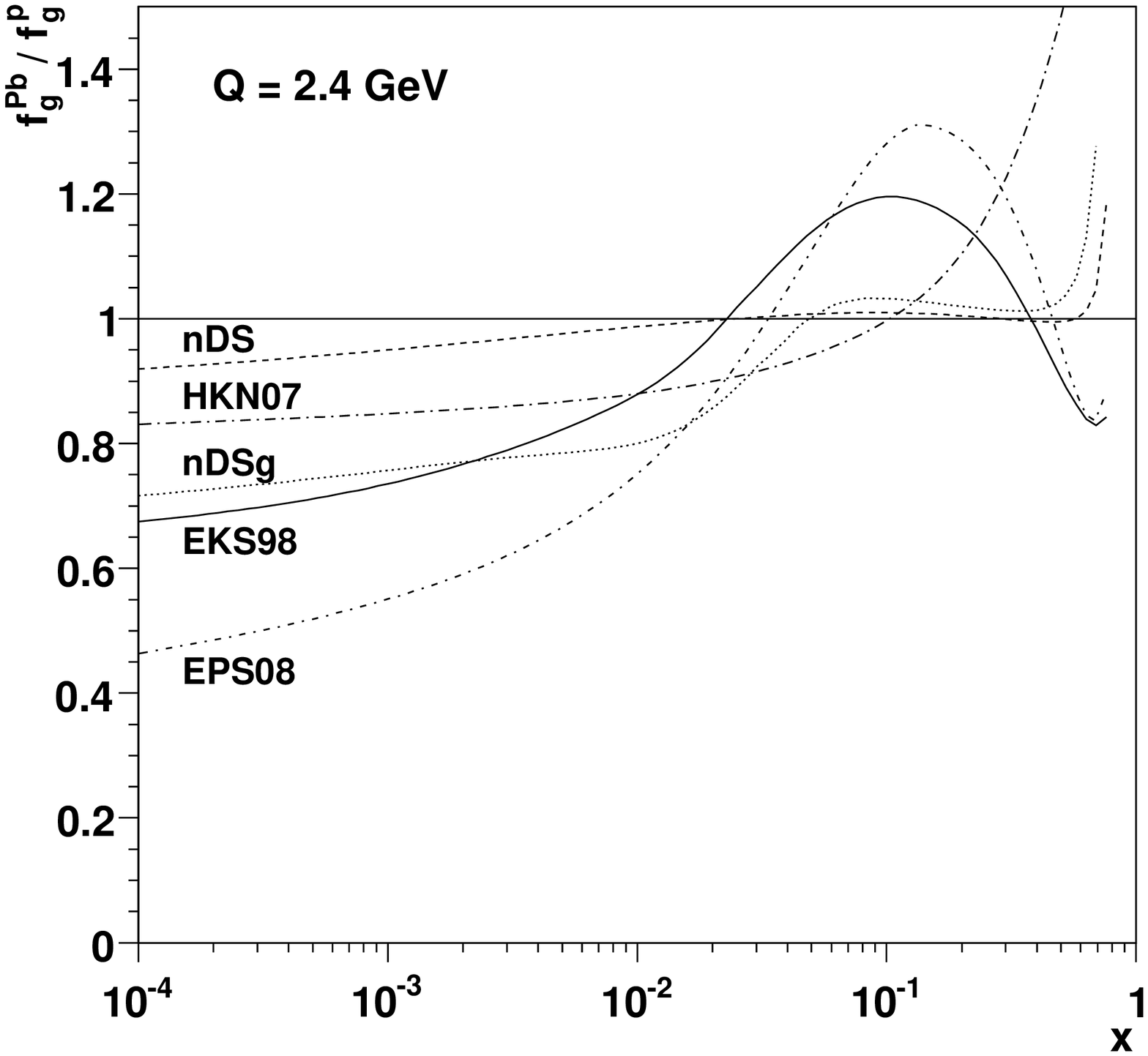}}
\caption{Ratio between the valence quark (left) and the gluon (right)
  distribution functions in a nucleon of a Pb nucleus and in a free
  proton, according to several models.}
\label{fig:shadowing}\end{figure}

Figure~\ref{fig:shadowing} shows several parametrisations of the
nuclear modification functions on a Pb nucleus for
valence quarks (left) and gluons (right), calculated for the
scale, $Q$, suitable for quarkonium production calculations with the
MRST2001\,LO~\cite{MRSTLO} or CTEQ61L~\cite{CTEQ6L} PDF sets.
Since the quark and antiquark distribution functions are directly
probed by the nuclear DIS and Drell-Yan data, their nuclear effects
are relatively well constrained and all parametrisations give similar
results.  The connection between the measurements and the nuclear
\emph{gluon} densities is much more indirect, however, relying on the
scale dependence of the $F_2$ structure function and on momentum sum
rules connecting the momentum distributions of gluons and quarks.  See
Ref.~\cite{eps08} (and references therein) for a recent review of the
problem and of the solutions explored so far.  This publication
presents the EPS08 model, which uses inclusive hadron production data
measured at forward rapidity by the BRAHMS experiment at RHIC to
further constrain the nuclear effects on the gluon densities.  Given
that those measurements might reflect other physical processes (such
as gluon saturation in a color-glass condensate), it is 
not at all clear that adding them in the derivation of the nuclear
PDFs is justified.  Therefore, in our study we prefer to place more
emphasis on the earlier EKS98~\cite{eks98} parametrisation, the first global
analysis of nuclear effects on the PDFs.  The nDS/nDSg~\cite{DS} and
the HKN07~\cite{hkn07} parametrisations represent alternative analyses, also
illustrated in Fig.~\ref{fig:shadowing}, available both at leading order
(used in our calculations) and at next-to-leading order.

Charmonium production at fixed-target energies probes $x$ values in
the ``antishadowing'' region, where the parton densities are enhanced
in the heavy nucleus.  Since \ccbar\ production is dominated by gluon
fusion, a good understanding of charmonium production in p-nucleus
collisions is presently hampered by the lack of detailed knowledge of
the nuclear gluon distributions, illustrated by the spread in the
curves shown in the right panel of Fig.~\ref{fig:shadowing}.

\begin{figure}[ht]\centering
\resizebox{0.48\textwidth}{!}{%
\includegraphics*{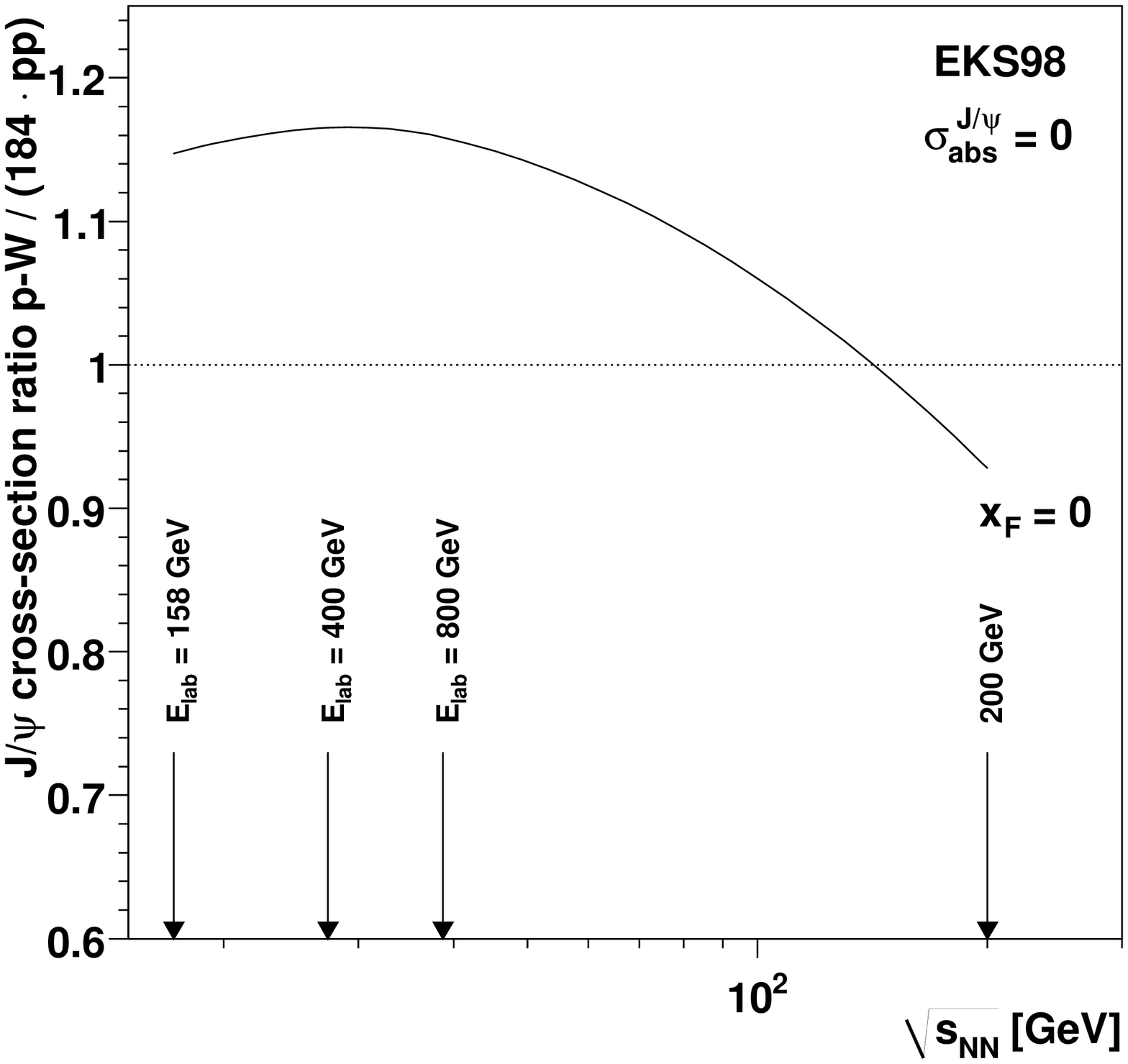}}
\resizebox{0.48\textwidth}{!}{%
\includegraphics*{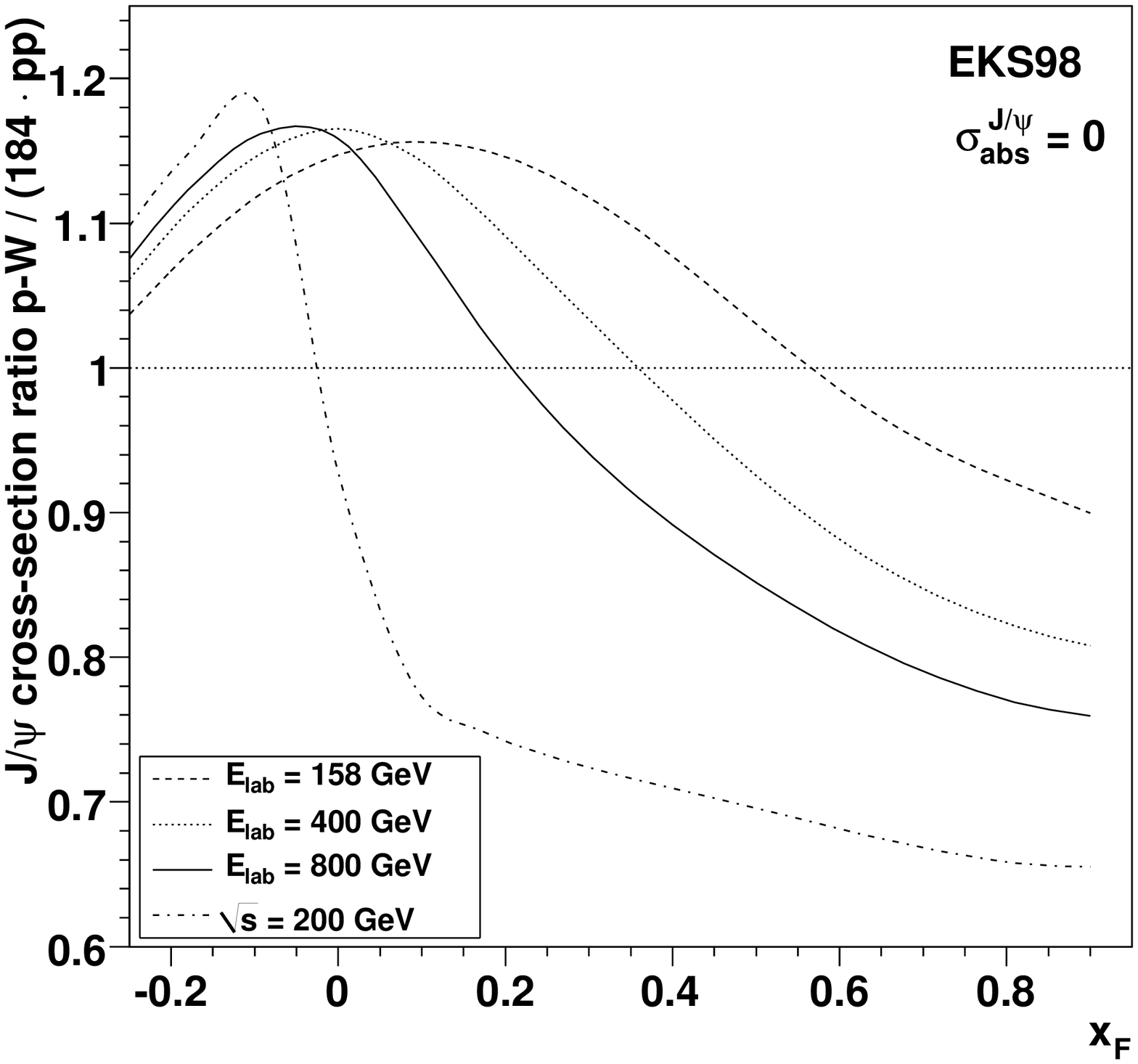}}
\caption{Changes induced by the nuclear modifications of the PDFs on
  the \ccbar\ production cross section per nucleon in \mbox{p-W}
  collisions, using EKS98: as a function of
  $\sqrt{s_{_{NN}}}$ at $x_{\rm F}=0$ (left) and as a function of
  $x_{\rm F}$ at four different energies
  (right).}
\label{fig:nucl-ccbar}\end{figure}

Figure~\ref{fig:nucl-ccbar} shows the influence on the \jpsi\
production cross section per nucleon (in the absence of any final
state absorption) of using the EKS98 \emph{nuclear} parton
distribution functions rather than those of the free proton.  The left
panel shows that, at fixed-target energies, midrapidity charmonium
production should be enhanced in p-Pb collisions with respect to the
linear extrapolation of the pp yields (see Refs.~\cite{CL-HKW, EKV03}
for more details).  The right panel shows that the ``antishadowing''
effect turns into a ``shadowing'' effect when charmonium
production at forward $x_{\rm F}$ is considered.

\subsection{Charmonium survival probabilities}
\label{sec:SurvProb}

In the framework of the Glauber model, described in detail in
Ref.~\cite{KLNS}, the probability that a given charmonium state,
generically represented by $\psi$, produced in a p-A collision,
traverses the target nucleus unbroken by interactions with the nuclear
matter, can be calculated as
\begin{equation}
S^\psi_{\rm abs} = \frac{\sigma^\psi_{\rm pA}}{A\, \sigma^\psi_{\rm pN}} =
\frac{1}{A} \, \int \dd^2b \, \int_{-\infty}^{\infty}\, \dd z ~
\rho_A(b,z) ~ S^\psi_{\rm abs}(b,z) \quad , \\
\label{sigfull}
\end{equation}
with
\begin{equation}
S^\psi_{\rm abs}(b,z) = \exp \left\{
- \int_z^{\infty} \dd z^{\prime} ~ \rho_A(b,z^{\prime})
~ \sigma^\psi_{\rm abs} \right\} \quad ,
\end{equation}
where $b$ is the impact parameter of the collision (transverse
distance between the flight path of the incident proton and the centre
of the nucleus) and $z$ is the $c \overline c$ production point along
the beam axis.  This ``survival probability'' depends essentially on
the nuclear density profiles, $\rho_A$, and on the charmonium break-up
cross section, $\sigma^\psi_{\rm abs}$.  In our calculations we used
Woods-Saxon density profiles with the parameters given in
Ref.~\cite{Jager}.  No charmonium absorption has been considered in
Hydrogen nuclei.

\begin{figure}[ht]\centering
\resizebox{0.5\textwidth}{!}{%
\includegraphics*{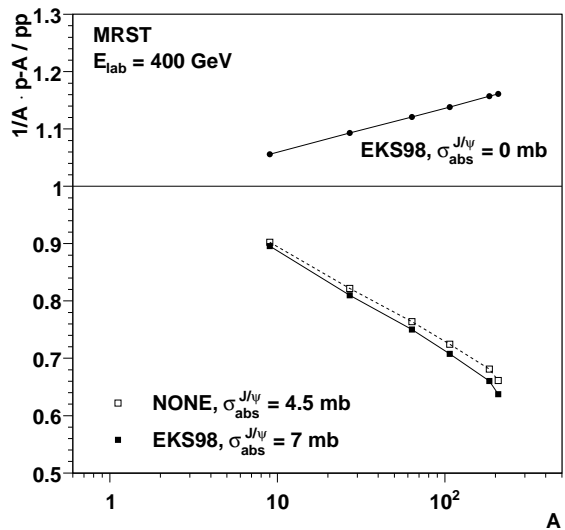}}
\caption{Illustration of the interplay between nuclear modifications
  of the parton densities and final-state charmonium absorption.}
\label{fig:ise-fse}\end{figure}

As we will see in detail later on, the absorption cross section
crucially depends on whether the PDFs are taken to be those of a free
nucleon or those of a nucleon in a nucleus.  For instance, at SPS
energies we obtain the same \jpsi\ nuclear absorption pattern using
proton PDFs and $\sigma_{\rm abs}^{\rm J/\psi} = 4.5$~mb as we do
using EKS98 nuclear PDFs and $\sigma_{\rm abs}^{\rm J/\psi} = 7$~mb,
as shown in Fig.~\ref{fig:ise-fse}.  If the \emph{enhancement} of the
per-nucleon charmonium production cross section caused by
initial-state antishadowing is ignored, a weaker \emph{final-state}
``effective'' absorption is sufficient to obtain the same result.

\section{Measured charmonium production cross sections}
\label{sec:data}

Table~\ref{tab:exp} lists the experiments that measured the charmonium
production cross sections, or ratios, used in our analysis.
All experiments detected the \jpsi\ through its decay to dimuons or
dielectrons, with good dilepton mass resolutions and small
backgrounds.

\begin{table}[ht]\centering
\caption{Basic features of the experiments providing the charmonium
production cross sections (or ratios) considered in the present study.}
\label{tab:exp}
\vglue2mm
\begin{tabular}{lclc} \hline
Experiment & E$_{\rm lab}$~[GeV] & Collision systems & Phase space\\ \hline
NA3~\cite{NA3} \rule{0pt}{0.5cm} & 200 & p-H,\,Pt & $0.0 < x_{\rm F} < 0.7$ \\
NA50~\cite{NA50-400} & 400 & p-Be,\,Al,\,Cu,\,Ag,\,W,\,Pb & $-0.425 < y_{\rm cms} < 0.575$ \\
NA50~\cite{NA50-450} & 450 & p-Be,\,Al,\,Cu,\,Ag,\,W & $-0.50 < y_{\rm cms} < 0.50$ \\
E866~\cite{E866}     & 800 & p-Be,\,W & $-0.10 < x_{\rm F} < 0.93$\\
\mbox{HERA-B}~\cite{HERA-B} & 920 & p-C,\,W & $-0.34 < x_{\rm F} < 0.14$ \\ \hline \hline
Experiment \rule{0pt}{0.5cm} & \ssnn~[GeV] & Collision systems & Phase space\\ \hline
PHENIX~\cite{PHENIX}  \rule{0pt}{0.5cm} & 200 
& \multicolumn{2}{c}{pp, d-Au \hfill $|y_{\rm cms}|<0.35$, $1.2<|y_{\rm cms}|<2.2$} \\ \hline
\end{tabular}\end{table}
\begin{table}[ht]\centering
  \caption{The \jpsi\ production cross sections, times branching ratio
    into dimuons, measured by NA50 at 400 and 450~GeV.  The errors
    include statistical and target-to-target systematic uncertainties,
    added in quadrature.}
\label{tab:NA50}
\vglue2mm
\begin{tabular}{cccc} \hline
\rule{0pt}{0.4cm} & \multicolumn{3}{c}{$B \times \sigma^{\rm J/\psi}\,/\,A$ [nb/nucleon]}\\
 & NA50-400 & NA50-450 ``LI'' & NA50-450 ``HI'' \\ \hline
Be & $4.717\pm0.10$ & $5.27\pm0.23$ & $5.11\pm0.18$\rule{0pt}{0.5cm}\\
Al & $4.417\pm0.10$ & $5.14\pm0.21$ & $4.88\pm0.23$ \\
Cu & $4.280\pm0.09$ & $4.97\pm0.22$ & $4.74\pm0.18$ \\       
Ag & $3.994\pm0.09$ & $4.52\pm0.20$ & $4.45\pm0.15$ \\       
W  & $3.791\pm0.08$ & $4.17\pm0.37$ & $4.05\pm0.15$ \\       
Pb & $3.715\pm0.08$ &        &  \\ \hline
\end{tabular}\end{table}

Since our study focuses on the nuclear dependence of the quarkonium
production cross section, it is preferable to have several target
nuclei at each energy.  
NA50 collected data with five or six different target materials, from
Be to Pb, providing a more detailed quarkonium absorption pattern as a
function of the size of the target nucleus than the experiments which
only used two targets.
The \jpsi\ production cross sections measured by NA50 in p-nucleus
collisions, at 400 and 450~GeV, in the centre-of-mass rapidity windows
$-0.425< y_{\rm cms} <0.575$ and $|y_{\rm cms}| < 0.5$, respectively,
are collected in Table~\ref{tab:NA50}.  The 450~GeV values correspond to
two statistically independent data sets, collected with ``low'' (LI)
and ``high'' (HI) intensity proton beams.  The 400~GeV errors are
dominated by a target-dependent relative systematic uncertainty of
2.1\,\%.  An extra global systematic uncertainty of 3\,\%, due to
common normalisation uncertainties, is not included because it does
not affect the evaluation of the nuclear dependence.  This is because
all the 400~GeV p-A data sets were collected in the same week,
changing the target exposed to the beam roughly every hour, using a
rotating target holder.  On the contrary, the 450~GeV data sets were
collected in different running periods, over five years, and have
independent normalisation uncertainties (due to beam counting, trigger
efficiencies, etc).  Therefore, the quoted 450~GeV errors reflect the
total uncertainties (added in quadrature).

The NA50 450~GeV cross sections were also reported in four equidistant
\xf\ bins~\cite{NA50-450}.  However, given their large uncertainties
and small \xf\ coverage, they do not really provide extra information
with respect to the integrated values, re-analysed in
Ref.~\cite{NA50-400} to ensure consistency with the 400~GeV data
analysis.
Since NA50 reported absolute production cross sections (in nb) for
each p-nucleus system, we extract the \jpsi\ nuclear absorption cross
section, \sabsjpsi, together with a $\sigma_0$ normalisation factor.
In the case of the other experiments we used a one-parameter fit to
extract \sabsjpsi\ from the \emph{ratios} between the per-nucleon
cross sections obtained with ``heavy'' and ``light'' nuclear targets.
The \mbox{HERA-B} and E866 data samples cover a relatively large range
in \xf\ and are very important for a differential study of the cold
nuclear matter effects.  Table~\ref{tab:E866HeraB} gives the E866
\jpsi\ W/Be ratios (see Ref.~\cite{E866} for the $x_{\rm F}>0.2$
values) and the \mbox{HERA-B} W/C ratios~\cite{HERA-B} (derived from
the exponent $\alpha$ using Eq.~\ref{eq:alpha}) which we have used in
our study.  The errors represent statistical and point-to-point
systematic uncertainties, added in quadrature.  The global
normalisation errors of 3\,\% (E866) and 4\,\% (HERA-B, obtained from
a 1.5\,\% uncertainty on $\alpha$) are not included in
Table~\ref{tab:E866HeraB} but must be considered when comparing
different experiments.  E866 also measured Fe/Be ratios but only for
$x_{\rm F}>0.2$.

\begin{table}[ht]\centering
  \caption{\jpsi\ cross section ratios measured by E866 and
    \mbox{HERA-B}, without including global errors (3\,\% and 4\,\%,
    respectively).}
\label{tab:E866HeraB}
\vglue2mm
\begin{tabular}{cccccc} \hline
\multicolumn{3}{c}{E866} & \multicolumn{3}{c}{HERA-B} \\
$x_{\rm F}$ range & $\langle x_{\rm F} \rangle$ & W\,/\,Be ratio &
$x_{\rm F}$ range & $\langle x_{\rm F} \rangle$ & W\,/\,C ratio \\ \hline
$-$0.10~/\,$-$0.05              & $-$0.0652 & $0.8929 \pm 0.0184$ &$-$0.34~/\,$-$0.26 & $-$0.285 & $1.105 \pm 0.158$\rule{0pt}{0.5cm}\\
$-$0.05~/\,\rule{10pt}{0pt}0.00 & $-$0.0188 & $0.8682 \pm 0.0084$ &$-$0.26~/\,$-$0.22 & $-$0.237 & $1.034 \pm 0.096$\\       
\rule{10pt}{0pt}0.00~/\,$+$0.05 & $+$0.0269 & $0.8720 \pm 0.0060$ &$-$0.22~/\,$-$0.18 & $-$0.197 & $1.090 \pm 0.063$\\       
$+$0.05~/\,$+$0.10              & $+$0.0747 & $0.8739 \pm 0.0057$ &$-$0.18~/\,$-$0.14 & $-$0.158 & $1.043 \pm 0.042$\\       
$+$0.10~/\,$+$0.15              & $+$0.1235 & $0.8652 \pm 0.0067$ &$-$0.14~/\,$-$0.10 & $-$0.118 & $0.986 \pm 0.030$\\       
$+$0.15~/\,$+$0.20              & $+$0.1729 & $0.8725 \pm 0.0100$ &$-$0.10~/\,$-$0.06 & $-$0.079 & $0.943 \pm 0.022$\\       
              & & &$-$0.06~/\,$-$0.02 & $-$0.040 & $0.915 \pm 0.021$\\       
              & & &$-$0.02~/\,$+$0.02 & $-$0.002 & $0.916 \pm 0.025$\\       
              & & &$+$0.02~/\,$+$0.06 & $+$0.037 & $0.902 \pm 0.036$\\       
              & & &$+$0.06~/\,$+$0.14 & $+$0.075 & $0.866 \pm 0.063$\\
%
\hline
\end{tabular}
\end{table}
\begin{table}[ht]\centering
  \caption{\jpsi\ cross section ratios measured by NA3 and PHENIX,
    without including global errors (estimated to be 3\,\% and 11\,\%, 
    respectively).}
\label{tab:NA3Phenix}
\vglue2mm
\begin{tabular}{cccc} \hline
\multicolumn{2}{c}{NA3} & \multicolumn{2}{c}{PHENIX} \\
$x_{\rm F}$ range & H\,/\,Pt ratio & $y_{\rm cms}$ range & d-Au\,/\,pp ratio \\ \hline
$0.0~/~0.1$ & $1.27 \pm 0.07$ & $-$2.2~/\,$-$1.7   & $0.95 \pm 0.23$\rule{0pt}{0.5cm}\\
$0.1~/~0.2$ & $1.40 \pm 0.06$ & $-$1.7~/\,$-$1.2   & $0.90 \pm 0.21$ \\
$0.2~/~0.3$ & $1.34 \pm 0.07$ & $-$0.35~/\,$+$0.35 & $0.85 \pm 0.17$ \\
$0.3~/~0.4$ & $1.36 \pm 0.12$ & $+$1.2~/\,$+$1.7   & $0.68 \pm 0.13$ \\
$0.4~/~0.5$ & $1.75 \pm 0.22$ & $+$1.7~/\,$+$2.2   & $0.59 \pm 0.12$ \\
$0.5~/~0.6$ & $2.62 \pm 0.52$ & & \\
$0.6~/~0.7$ & $3.58 \pm 1.81$ & & \\ \hline
\end{tabular}
\end{table}

Table~\ref{tab:NA3Phenix} gives the NA3 H/Pt ratios, for several \xf\
bins, as extracted ``by eye'' from Fig.~2 of Ref.~\cite{NA3}.  A
common systematic error of 3\,\%~\cite{PhC} is not included.
NA3 used two targets: protons and platinum.
When comparing the NA3 values to the other results, we should keep in
mind that a proton is not exactly a ``nucleus''.
We also analysed the presently available PHENIX \mbox{d-Au/pp} ratios
in several rapidity bins~\cite{PHENIX}, as collected in
Table~\ref{tab:NA3Phenix}.  These ratios have essentially no impact in
the results of our study, given their large uncertainties.

It might be worth mentioning that we have not included in our study
the measurements reported by the NA38 and E772 experiments.  Later
analyses of those data sets, made in the framework of the
NA50~\cite{Goncalo,Goncalo-HP04} and E866~\cite{E866} experiments,
respectively, revealed that those early results were biased, because
of wrongly evaluated reconstruction efficiencies (NA38) or acceptances
(E772).

\section{Analysis of charmonium nuclear absorption}
\label{sec:analysis}

\subsection{Cold nuclear matter effects}

\begin{figure}[htb]\centering
\resizebox{0.48\textwidth}{!}{%
\includegraphics*{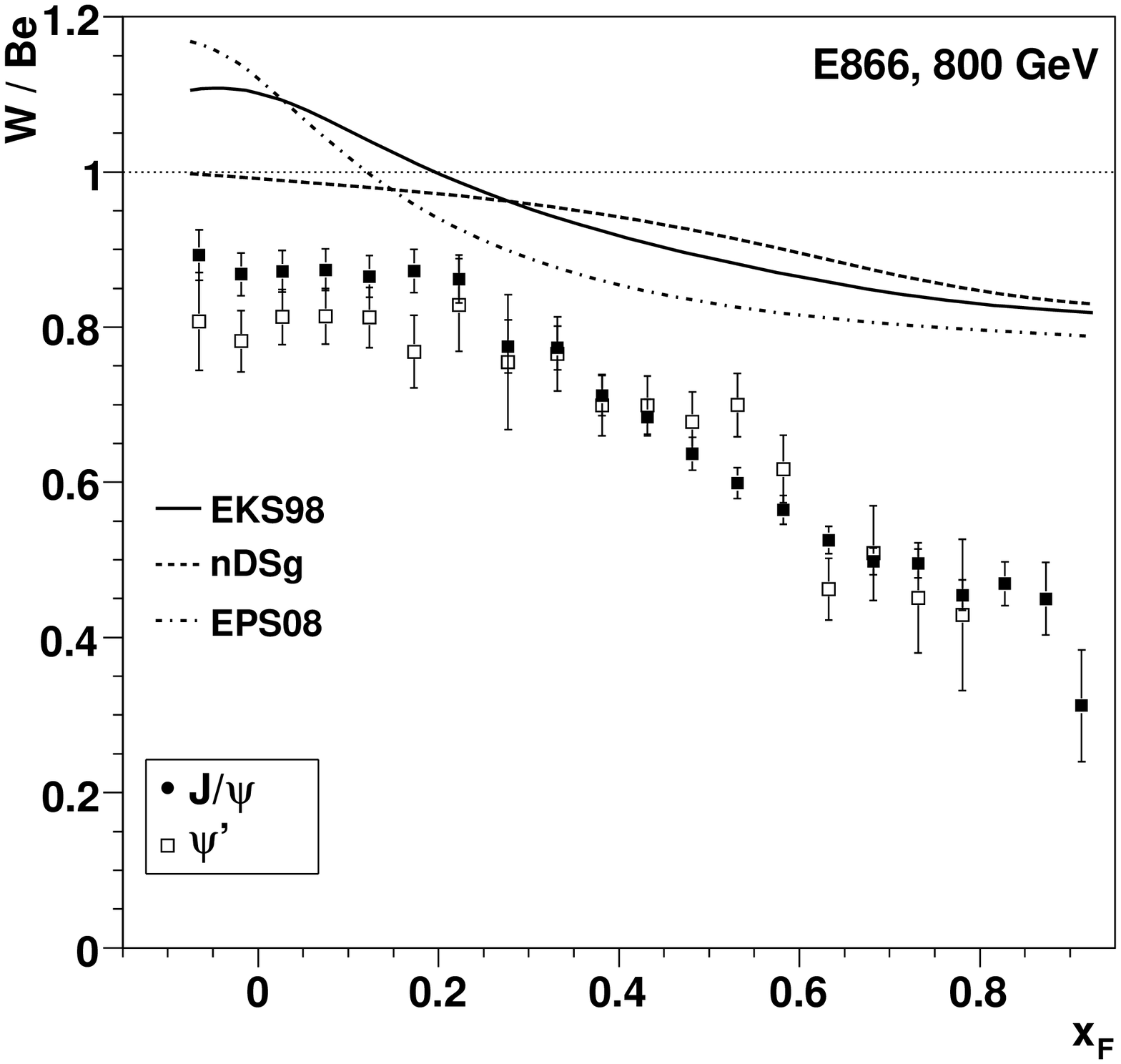}}
\resizebox{0.48\textwidth}{!}{%
\includegraphics*{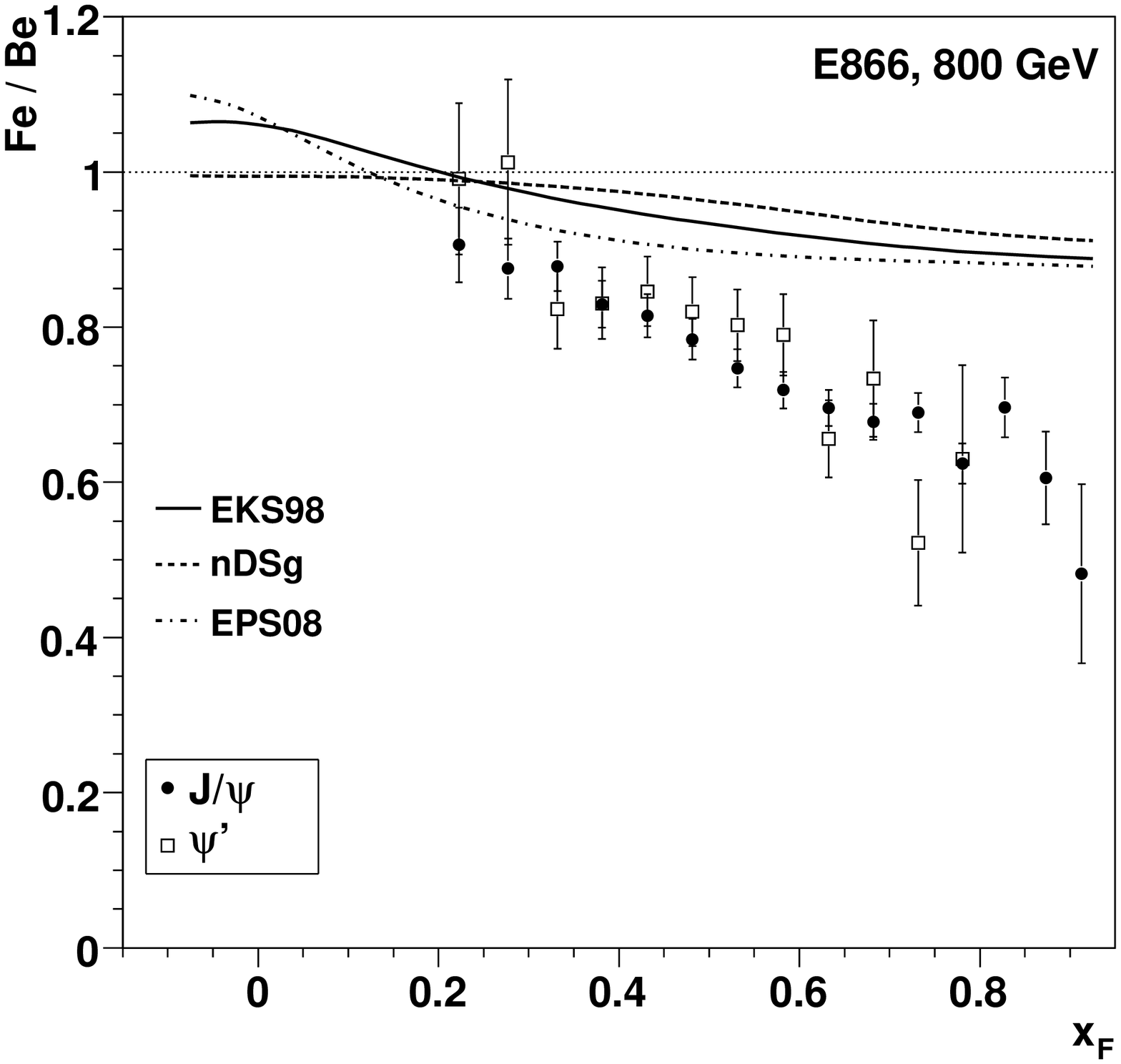}}
\caption{The \xf\ dependence of the \jpsi\ and \psip\ production
  ratios, W/Be (left) and Fe/Be (right), per nucleon, measured by E866
  at $E_{\rm lab} = 800$~GeV (points) and calculated assuming only
  nuclear modifications of the parton densities (curves).}
\label{fig:ratioE866_xF}
\end{figure}

Now that we have reviewed the available \jpsi\ production measurements
in proton-nucleus collisions, we can study them in view of deriving
the so-called normal nuclear absorption, described in the
framework of the Glauber formalism and quantified through the ``\jpsi\
break-up cross section'', \sabsjpsi, introduced in
Section~\ref{sec:SurvProb}.
Before proceeding, however, we note that this is not the only
mechanism affecting the per-nucleon production cross sections in
p-nucleus collisions.  As mentioned in Sections~\ref{sec:cem}
and~\ref{sec:nPDFs}, the quarkonium production cross sections
crucially depend on the parton densities (particularly the gluon
densities), which are significantly affected by nuclear modifications.
These modifications have been taken into account in our calculations,
employing several parametrisations.  Other nuclear effects, such as
energy loss, formation times, etc., are likely to be present and would
need to be considered in a detailed study of all aspects of quarkonium
``cold nuclear effects''.  In particular, the nuclear charmonium
production yields measured by E866 are clearly more suppressed at
forward \xf\ than at $x_{\rm F}=0$, as observed in the W/Be and Fe/Be
ratios shown in Fig.~\ref{fig:ratioE866_xF}.  
There is a remarkable change in the suppression pattern at $x_{\rm F}
\approx 0.25$, from a relatively flat region around $x_{\rm F} \approx
0$, where the \psip\ is more absorbed than the \jpsi, to a forward
region where both states show the same strong decrease with increasing
$x_{\rm F}$.

The curves in Fig.~\ref{fig:ratioE866_xF} show the expected trends
when only considering the nuclear modifications on the parton
distribution functions, without any final state absorption.  It is
worth noting the antishadowing enhancement expected in the EKS98
and EPS08 models at $x_{\rm F} \approx 0$.

In the present work, we concentrate on the ``midrapidity region'' and
neglect nuclear effects other than initial-state modifications of
the parton densities and final-state charmonium absorption, calculated
as described in Sections~\ref{sec:cem} and~\ref{sec:nPDFs}.  As in
most previous studies of charmonium absorption in nuclear matter, we
treat the \jpsi\ as a single meson passing through the nuclear medium.
However, significant fractions of the \jpsi\ yield observed in
elementary collisions are due to $\psi^\prime$ and $\chi_c$ decays,
8.1\,$\pm$\,0.3~\% and 25\,$\pm$\,5~\%, respectively, as recently
determined from charmonium hadroproduction at $x_{\rm F} \approx
0$~\cite{feeddown}.

\subsection{\boldmath Extraction of \sabsjpsi\ from the measurements}

From the equations in Section~\ref{sec:theory}, we can see that the
per-nucleon heavy-to-light target ratio of the charmonium production
cross sections decreases as a function of the break-up cross section.
This dependence has been calculated respecting the conditions of each
measurement, considering the kinematics window, the collision energy
and the nuclear matter densities.  The calculations were performed
using three sets of proton parton distribution functions:
GRV\,LO\,98~\cite{GRV98}; MRST2001\,LO~\cite{MRSTLO}; and
CTEQ61L~\cite{CTEQ6L}.  In addition to a control calculation with no
nuclear modifications of the PDFs, labelled NONE, we employ four
models of the nuclear modifications: EKS98~\cite{eks98};
nDSg~\cite{DS}; nDS~\cite{DS}; and EPS08~\cite{eps08}.

\begin{figure}[ht]\centering
\resizebox{0.5\textwidth}{!}{%
\includegraphics*{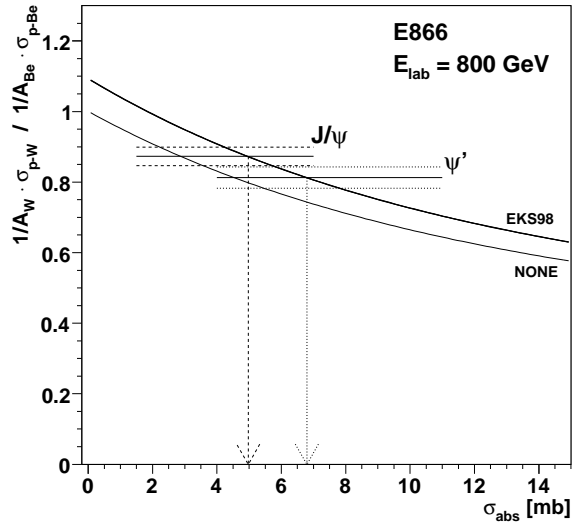}}
\caption{Illustration of extracting \sabs\ for \jpsi\ and \psip\
  production from the E866 \mbox{W/Be} ratio with and without the
  EKS98 modifications of the PDFs.}
\label{fig:extractSigma_E866}
\end{figure}

The rather straightforward extraction of \sabs\ for each measurement
is illustrated in Fig.~\ref{fig:extractSigma_E866}, using the E866
\jpsi\ and \psip\ \mbox{W/Be} ratios in the range $-0.1 <x_{\rm F} <
0.1$.  The PDFs effectively cancel in the calculations of the
heavy/light ratios, even when nuclear effects are considered, since
charmonium production is dominated by $gg$ fusion.
Given that the NA50 data were collected with five or six
different targets, we performed a global two-parameter fit to all the
p-A production cross sections within each data set (400, 450-LI and
450-HI).

\begin{table}[ht]\centering
  \caption{The \jpsi\ break-up cross sections, \sabsjpsi, obtained in
    the Glauber framework described in Section~\ref{sec:theory}. The
    NA50 400 and 450~GeV values correspond to the rapidity windows
    $-0.425<y_{\rm cms}<0.575$ and $|y_{\rm cms}| < 0.5$,
    respectively.  Our study uses averages of the two 450~GeV sets.}
\label{tab:sigmaAbs-jPsi}
\vglue2mm
\begin{tabular}{lccccc} \hline
\rule{0pt}{0.5cm}
Exp. \hfill \xf~~ & \multicolumn{5}{c}{\sabsjpsi\ [mb]}\rule{0pt}{0.5cm}\\ 
               &    NONE     &    nDS      &    nDSg     &    EKS98    &    EPS08     \\ \hline
NA3 \hfill 0.05 & $3.77\pm0.98$ & $3.94\pm0.99$ & $4.27\pm1.00$ & $5.79\pm1.07$ & $7.00\pm1.12$\rule{0pt}{0.5cm} \\
    \hfill 0.15 & $5.35\pm0.88$ & $5.46\pm0.88$ & $5.85\pm0.89$ & $7.38\pm0.95$ & $8.15\pm0.98$ \\
    \hfill 0.25 & $4.66\pm0.98$ & $4.63\pm0.98$ & $5.01\pm0.99$ & $6.18\pm1.04$ & $6.38\pm1.05$ \\
    \hfill 0.35 & $4.96\,^{+\,1.51}_{-\,1.56}$ & $4.71\,^{+\,1.49}_{-\,1.54}$ & $5.07\,^{+\,1.51}_{-\,1.56}$ & $5.81\,^{+\,1.56}_{-\,1.61}$ & $5.62\,^{+\,1.55}_{-\,1.60}$\\[1mm] \hline
NA50-400 \hfill  & $4.83\pm0.63$ & $4.74\pm0.62$ & $4.73\pm0.62$ & $7.01\pm0.70$ & $7.98\pm0.74$\rule{0pt}{0.5cm} \\
~~ 450-LI \hfill & $4.51\pm1.58$ & $4.39\pm1.58$ & $4.39\pm1.58$ & $6.89\pm1.76$ & $7.93\pm1.83$\\
~~ 450-HI \hfill & $4.82\pm1.10$ & $4.71\pm1.09$ & $4.71\pm1.09$ & $7.17\pm1.22$ & $8.21\pm1.28$\\ \hline
E866 
\hfill $-$0.0652 & $2.37\,^{+\,0.83}_{-\,0.77}$ & $2.32\,^{+\,0.83}_{-\,0.77}$ & $3.01\,^{+\,0.85}_{-\,0.79}$ & $4.67\,^{+\,0.92}_{-\,0.85}$ & $6.06\,^{+\,0.98}_{-\,0.90}$\rule{0pt}{0.5cm} \\
\hfill $-$0.0188 & $3.00\,^{+\,0.73}_{-\,0.69}$ & $2.85\,^{+\,0.73}_{-\,0.69}$ & $3.62\,^{+\,0.75}_{-\,0.71}$ & $5.39\,^{+\,0.82}_{-\,0.76}$ & $6.20\,^{+\,0.85}_{-\,0.79}$\\
\hfill $+$0.0269 & $2.90\,^{+\,0.71}_{-\,0.67}$ & $2.65\,^{+\,0.70}_{-\,0.66}$ & $3.27\,^{+\,0.72}_{-\,0.68}$ & $4.98\,^{+\,0.78}_{-\,0.73}$ & $5.03\,^{+\,0.78}_{-\,0.73}$\\
\hfill $+$0.0747 & $2.85\,^{+\,0.71}_{-\,0.67}$ & $2.50\,^{+\,0.70}_{-\,0.66}$ & $2.65\,^{+\,0.70}_{-\,0.66}$ & $4.36\,^{+\,0.76}_{-\,0.71}$ & $3.81\,^{+\,0.74}_{-\,0.70}$\\
\hfill $+$0.1235 & $3.07\,^{+\,0.72}_{-\,0.68}$ & $2.61\,^{+\,0.71}_{-\,0.67}$ & $2.13\,^{+\,0.69}_{-\,0.65}$ & $3.95\,^{+\,0.75}_{-\,0.71}$ & $2.98\,^{+\,0.72}_{-\,0.68}$\\
\hfill $+$0.1729 & $2.89\,^{+\,0.74}_{-\,0.70}$ & $2.31\,^{+\,0.73}_{-\,0.68}$ & $1.28\,^{+\,0.69}_{-\,0.65}$ & $3.13\,^{+\,0.75}_{-\,0.71}$ & $1.91\,^{+\,0.71}_{-\,0.67}$\\[1mm] \hline
\mbox{HERA-B} 
\hfill $-$0.158 &     ---       &     ---       &     ---       & $0.73\,^{+\,1.42}_{-\,0.73}$ & $2.23\,^{+\,1.52}_{-\,1.35}$\rule{0pt}{0.5cm} \\
\hfill $-$0.118 & $0.34\,^{+\,1.22}_{-\,0.34}$ & $0.42\,^{+\,1.22}_{-\,0.42}$ & $0.96\,^{+\,1.25}_{-\,0.96}$ & $2.34\,^{+\,1.33}_{-\,1.20}$ & $3.88\,^{+\,1.43}_{-\,1.28}$\\
\hfill $-$0.079 & $1.39\,^{+\,1.18}_{-\,1.08}$ & $1.38\,^{+\,1.18}_{-\,1.08}$ & $2.04\,^{+\,1.22}_{-\,1.11}$ & $3.68\,^{+\,1.32}_{-\,1.19}$ & $5.08\,^{+\,1.41}_{-\,1.26}$\\
\hfill $-$0.040 & $2.11\,^{+\,1.21}_{-\,1.10}$ & $1.99\,^{+\,1.20}_{-\,1.09}$ & $2.76\,^{+\,1.24}_{-\,1.13}$ & $4.53\,^{+\,1.36}_{-\,1.22}$ & $5.46\,^{+\,1.42}_{-\,1.27}$\\
\hfill $-$0.002 & $2.10\,^{+\,1.28}_{-\,1.15}$ & $1.85\,^{+\,1.26}_{-\,1.14}$ & $2.58\,^{+\,1.31}_{-\,1.18}$ & $4.32\,^{+\,1.42}_{-\,1.27}$ & $4.58\,^{+\,1.44}_{-\,1.29}$\\
\hfill $+$0.037 & $2.46\,^{+\,1.51}_{-\,1.34}$ & $2.09\,^{+\,1.49}_{-\,1.32}$ & $2.51\,^{+\,1.52}_{-\,1.35}$ & $4.28\,^{+\,1.65}_{-\,1.45}$ & $3.94\,^{+\,1.63}_{-\,1.43}$\\
\hfill $+$0.075 & $3.52\,^{+\,2.43}_{-\,2.02}$ & $2.96\,^{+\,2.36}_{-\,1.97}$ & $2.52\,^{+\,2.31}_{-\,1.93}$ & $4.58\,^{+\,2.56}_{-\,2.11}$ & $3.59\,^{+\,2.44}_{-\,2.02}$\\[1mm] \hline
\end{tabular}
\end{table}

Table~\ref{tab:sigmaAbs-jPsi} collects the \sabsjpsi\ values obtained
with the CTEQ61L PDFs, including the aforementioned nuclear
modifications.  Other PDFs give essentially the same results.  
The global uncertainties on the cross-section ratios were propagated
into the errors on \sabsjpsi\ using a Monte Carlo procedure; they are
common to all the \xf\ bins at the level of the ratios but not at the
level of \sabsjpsi.  This way, we can directly compare results
obtained by different experiments.
We can clearly see a correlation between the extracted \sabsjpsi\
values and the assumed initial-state nuclear effects on the PDFs
(already indicated in Fig.~\ref{fig:ise-fse}): a strong antishadowing
effect leads to a large \sabsjpsi.
The most backward \mbox{HERA-B} bins, $x_{\rm F} < -0.14$, where
$\alpha > 1$, do not require any final-state absorption unless strong
antishadowing effects, such as those in EKS98 and EPS08, enhance the
initial-state production rates in heavy targets.
Given the large uncertainties of the presently available PHENIX
ratios, we do not include the
corresponding \sabsjpsi\ values in Table~\ref{tab:sigmaAbs-jPsi}.

\subsection{\boldmath Dependence of \sabsjpsi\ on kinematics}

\begin{figure}[ht]\centering
\resizebox{0.48\textwidth}{!}{%
\includegraphics*{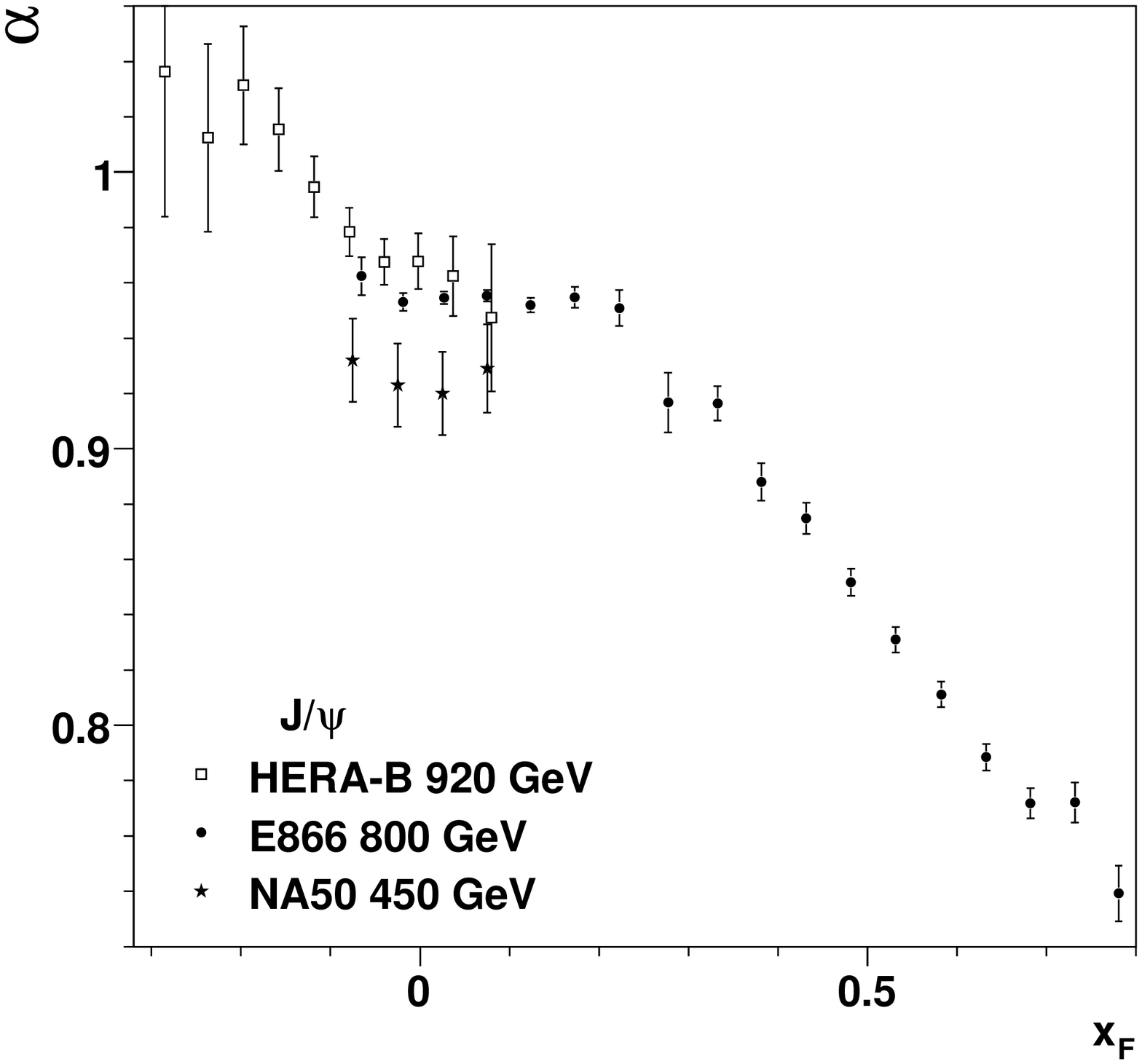}}
\resizebox{0.48\textwidth}{!}{%
\includegraphics*{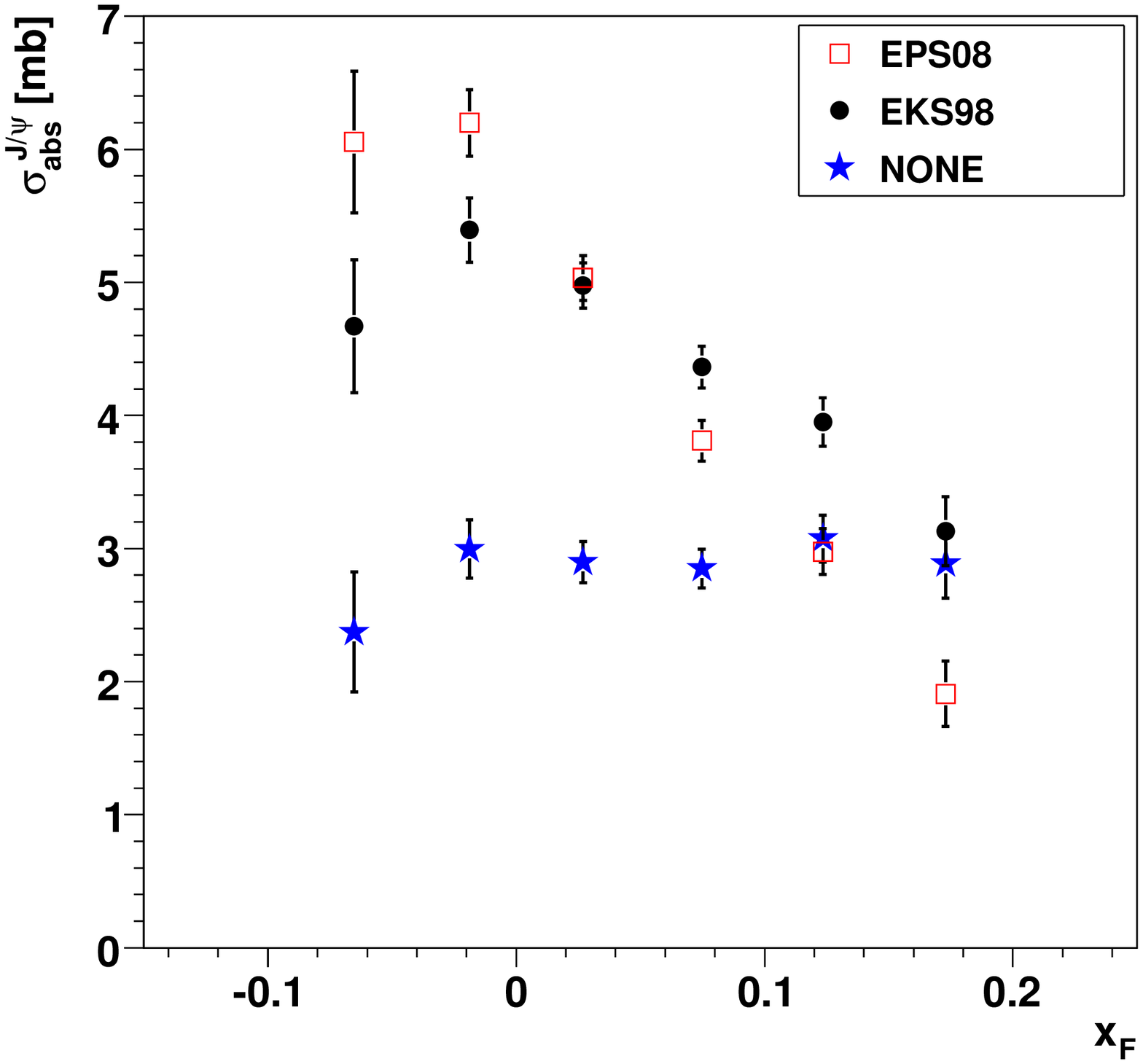}}
\caption{The nuclear dependence of the charmonium production cross
  sections measured by E866, expressed in terms of $\alpha$ (left) and
  \sabsjpsi\ (right), without the 3\,\% global uncertainty.  The left
  panel also includes the NA50 and HERA-B patterns.}
\label{fig:e866alpha}
\end{figure}

The ratio between the p-W and the p-Be \jpsi\ (and \psip) cross
sections measured by E866, shown in Fig.~\ref{fig:ratioE866_xF},
indicates the existence of additional charmonium absorption mechanisms at
forward \xf, perhaps related to formation time and energy loss
effects.  On the other hand, the measured pattern is remarkably flat
in the region $x_{\rm F} < 0.25$, an observation naturally interpreted
as meaning that the processes responsible for the strong absorption at
forward \xf\ are negligible at midrapidity.  This justifies confining
our present studies to the ``midrapidity region'', where the nuclear
effects are seemingly simpler and ``Glauber final-state charmonium
absorption'', as described in Section~\ref{sec:SurvProb}, may be
sufficient to describe most measurements.  Furthermore, the SPS
heavy-ion data samples have been collected close to midrapidity.
Obviously, the flat W/Be ratios at ``midrapidity'' directly translate
into an equally flat $\alpha(x_{\rm F})$, as shown in the left panel
of Fig.~\ref{fig:e866alpha}.

\begin{figure}[ht]\centering
\resizebox{0.48\textwidth}{!}{%
\includegraphics*{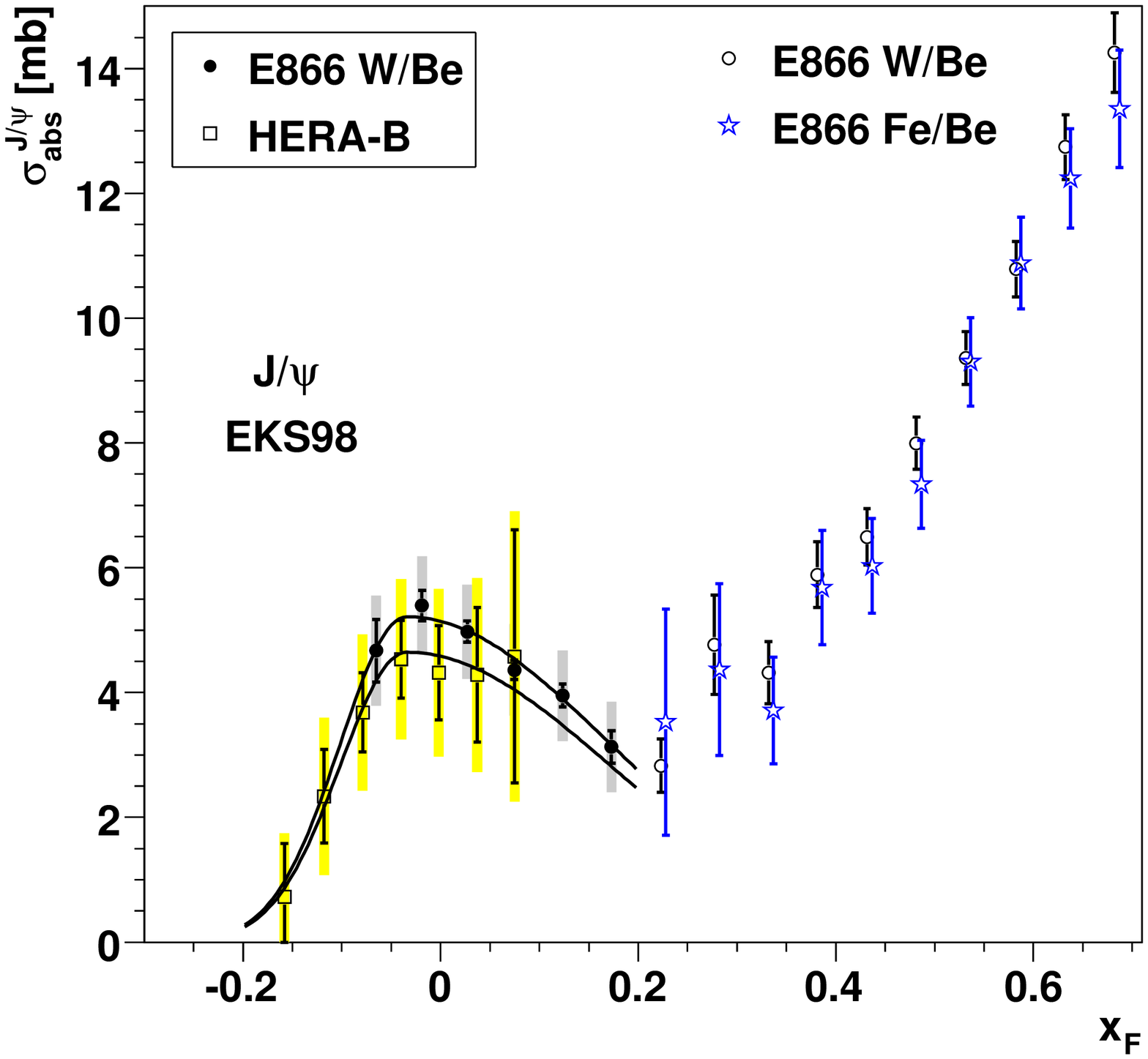}}
\resizebox{0.48\textwidth}{!}{%
\includegraphics*{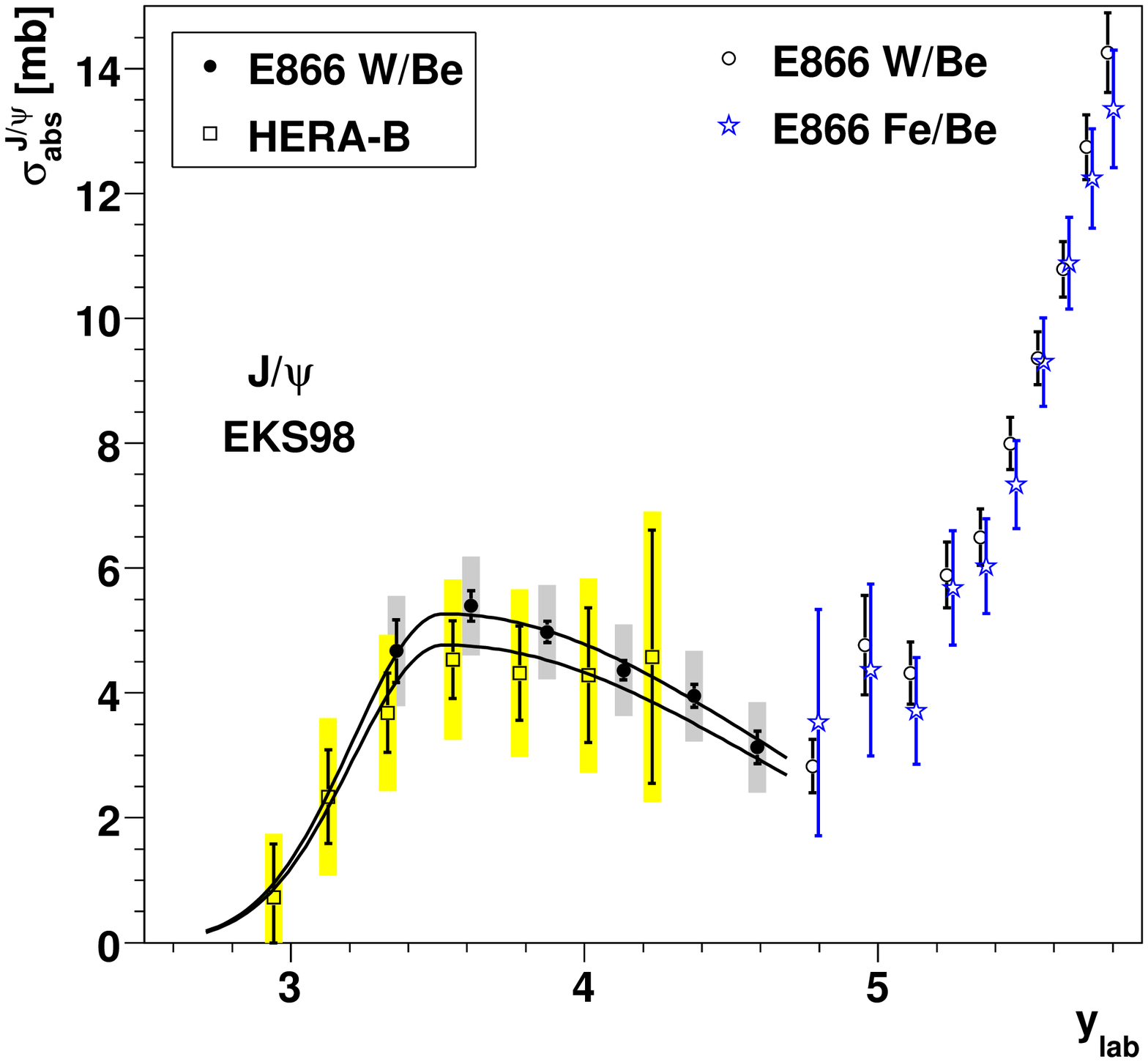}}
\caption{The extracted \sabsjpsi\ as a function of \xf\ (left) and
  $y_{\rm lab}$ (right), as derived from the E866
  and \mbox{HERA-B} data. The forward E866 data points (open circles
  and stars) are not included in the fitted curves.  The boxes
  represent the total errors.}
\label{fig:sigmaAbs_differential}
\end{figure}

Our calculations also give rather constant \sabsjpsi\ values in the
range $-0.1<x_{\rm F}<+0.2$, with $\sigma_{\rm abs}^{\rm J/\psi}\sim
3$~mb, when we neglect nuclear modifications of the PDFs, as shown
numerically in Table~\ref{tab:sigmaAbs-jPsi} and graphically in the
right panel of Fig.~\ref{fig:e866alpha}.  However, when we use nuclear
parton distributions with strong antishadowing, we see that \sabsjpsi\
is significantly larger at $x_{\rm F}=-0.1$ than at $+0.2$, (see
Fig.~\ref{fig:e866alpha}-right).  Another observation can be
made from the \xf\ dependence of the E866 data.  If we assume
that nuclear absorption effects on the \jpsi\ can be effectively
described by the Glauber formalism with a single \sabsjpsi\ (ignoring
formation times, feed-down contributions, energy loss, etc), we see a
rather striking \emph{increase} of \sabsjpsi\ at forward \xf,
explicitly shown in Fig.~\ref{fig:sigmaAbs_differential} but already
clear in Fig.~\ref{fig:ratioE866_xF} from the difference between the
calculations and the data.  The observation of a stronger absorption
at forward rapidity (or \xf) does not depend on the nuclear
modifications model used.

The \mbox{HERA-B} measurements also span a relatively large \xf\
range, covering a more backward window but having a sizable overlap
with E866 around $x_{\rm F} \approx 0$.
Figure~\ref{fig:sigmaAbs_differential} shows the E866 and
\mbox{HERA-B} \sabsjpsi\ values, obtained using the EKS98 nuclear
PDFs, as a function of $x_{\rm F}$ and laboratory rapidity, 
$y_{\rm lab}$, calculated from \xf\ accounting for the \jpsi\ $\langle
p_{\rm T}^2\rangle$ (which increases with \ssnn).  
The ``midrapidity'' ($x_{\rm F} < 0.25$ or $y_{\rm lab} < 5$) E866 and
\mbox{HERA-B} patterns can be empirically parametrised by the same
asymmetric Gaussian shape
but independent magnitudes.

Before continuing, we reiterate our original goal in light of what we have
observed so far.  We seek to determine and quantify possible changes
of the ``midrapidity \jpsi\ break-up cross section'' as a function of
collision energy.  More specifically, we want to find out whether it is
justified to analyse heavy-ion measurements at 158~GeV using the
\sabsjpsi\ values derived from proton-nucleus data collected at much
higher energies (400 and 450~GeV).  We have now seen that, in reality,
there is no single ``midrapidity'' \sabsjpsi\ value, unless we
neglect nuclear effects on the PDFs.  In particular,
the E866 data indicate that, for nDSg, EKS98 or EPS08, \sabsjpsi\
drops by a factor of 2 or 3 in the range $0 < x_{\rm F} < 0.25$,
corresponding to the centre-of-mass rapidity range $0 < y_{\rm cms} <
1$.  Such a non-negligible change in \sabsjpsi\ with the longitudinal
momentum of the \jpsi\ indicates that our quest is not an easy one.
There are several possible scenarios worth considering.  For example,
it could turn out that all data sets, regardless of collision energy,
exhibit the same \sabsjpsi\ dependence on $y_{\rm cms}$, in shape and
magnitude.  If so, we can determine the \sabsjpsi\ suitable for the
analysis of the heavy-ion data by integrating that universal function
in the range $0< y_{\rm cms} < 1$, the NA50 spectrometer window at
158~GeV.  This is likely to result in a different \sabsjpsi\ than that
obtained by integrating the same function in the 450~GeV window,
$|y_{\rm cms}|<0.5$.  On the other hand, if the magnitude of
\sabsjpsi\ is energy dependent, the change in the rapidity window
could, perhaps, partially compensate the change in energy, leading to
similar \sabsjpsi\ values at 158~GeV in $0 <y_{\rm cms}<1$ and at
450~GeV in $|y_{\rm cms}|<0.5$.  It is also possible that the
\emph{shape} of $\sigma_{\rm abs}^{\rm J/\psi}(y_{\rm cms})$ depends
on energy, in which case $y_{\rm cms}$ is not a suitable variable to
describe charmonium absorption.

\subsection{\boldmath Dependence of \sabsjpsi\ on the collision energy}

To evaluate which scenario best describes the measurements requires a
global survey of all available p-A results, obtained at different
energies and in different \xf\ or rapidity ranges.

\begin{figure}[ht!]\centering
\resizebox{0.48\textwidth}{!}{%
\includegraphics*{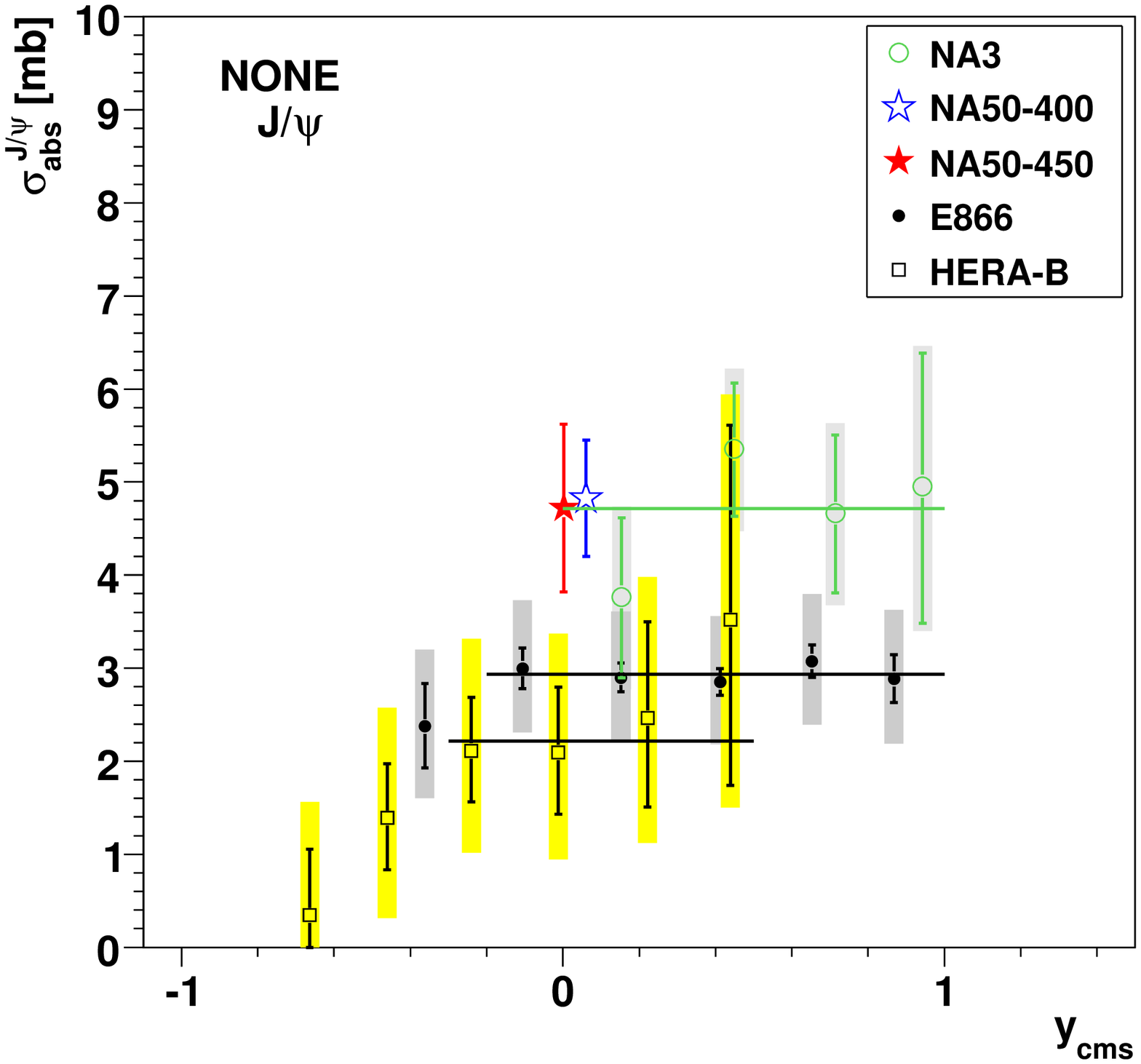}}
\resizebox{0.48\textwidth}{!}{%
\includegraphics*{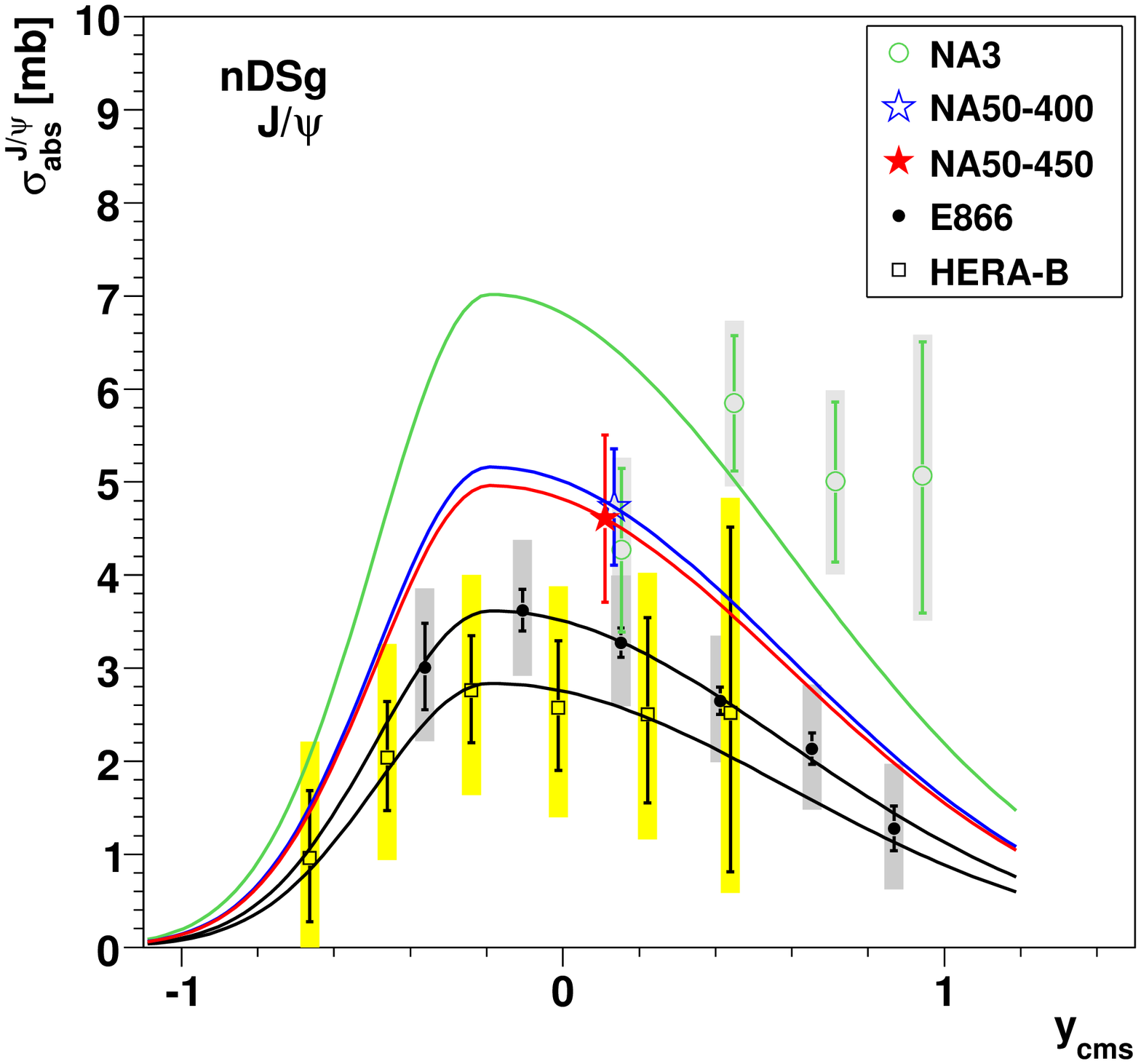}}
\resizebox{0.48\textwidth}{!}{%
\includegraphics*{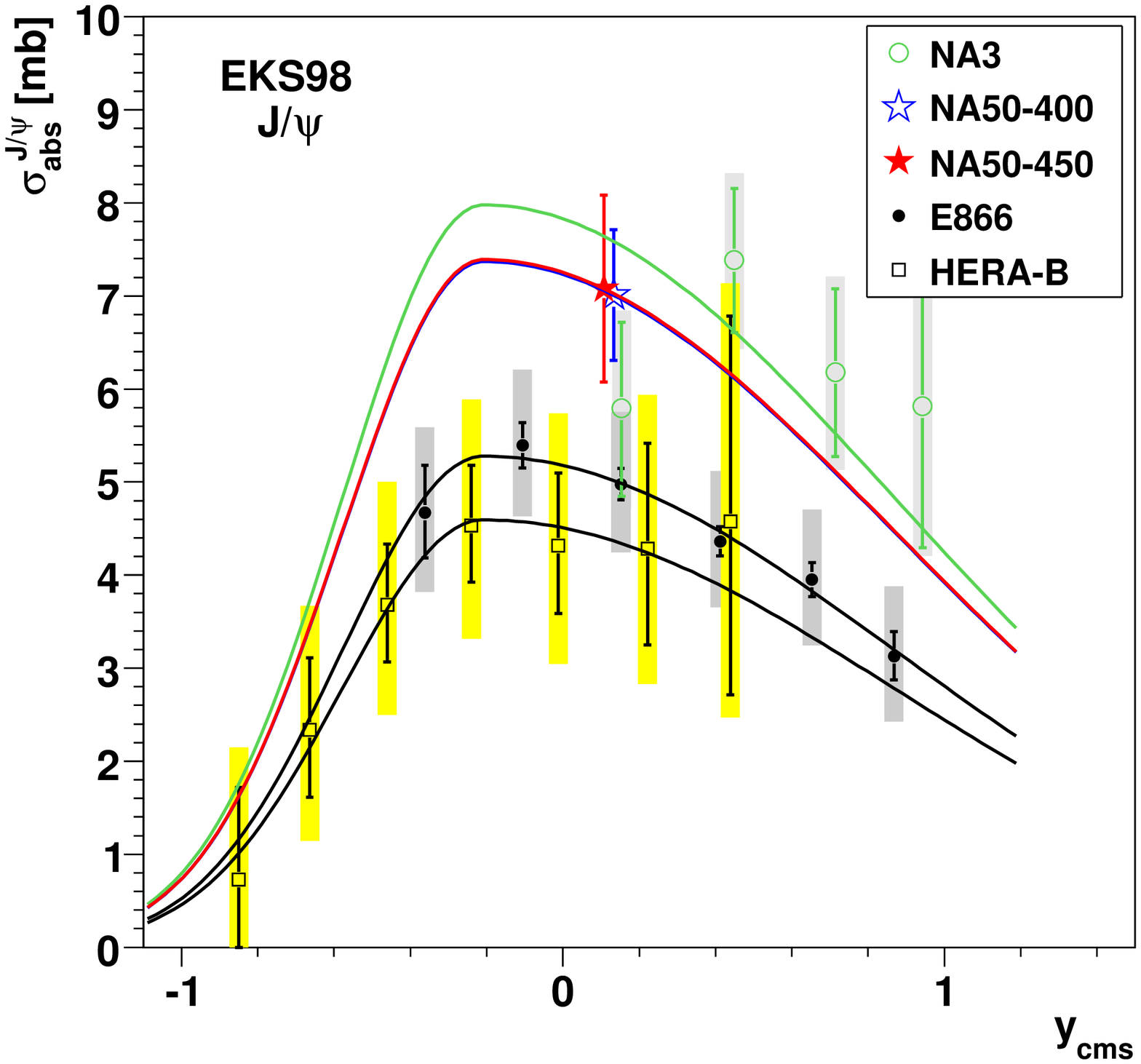}}
\resizebox{0.48\textwidth}{!}{%
\includegraphics*{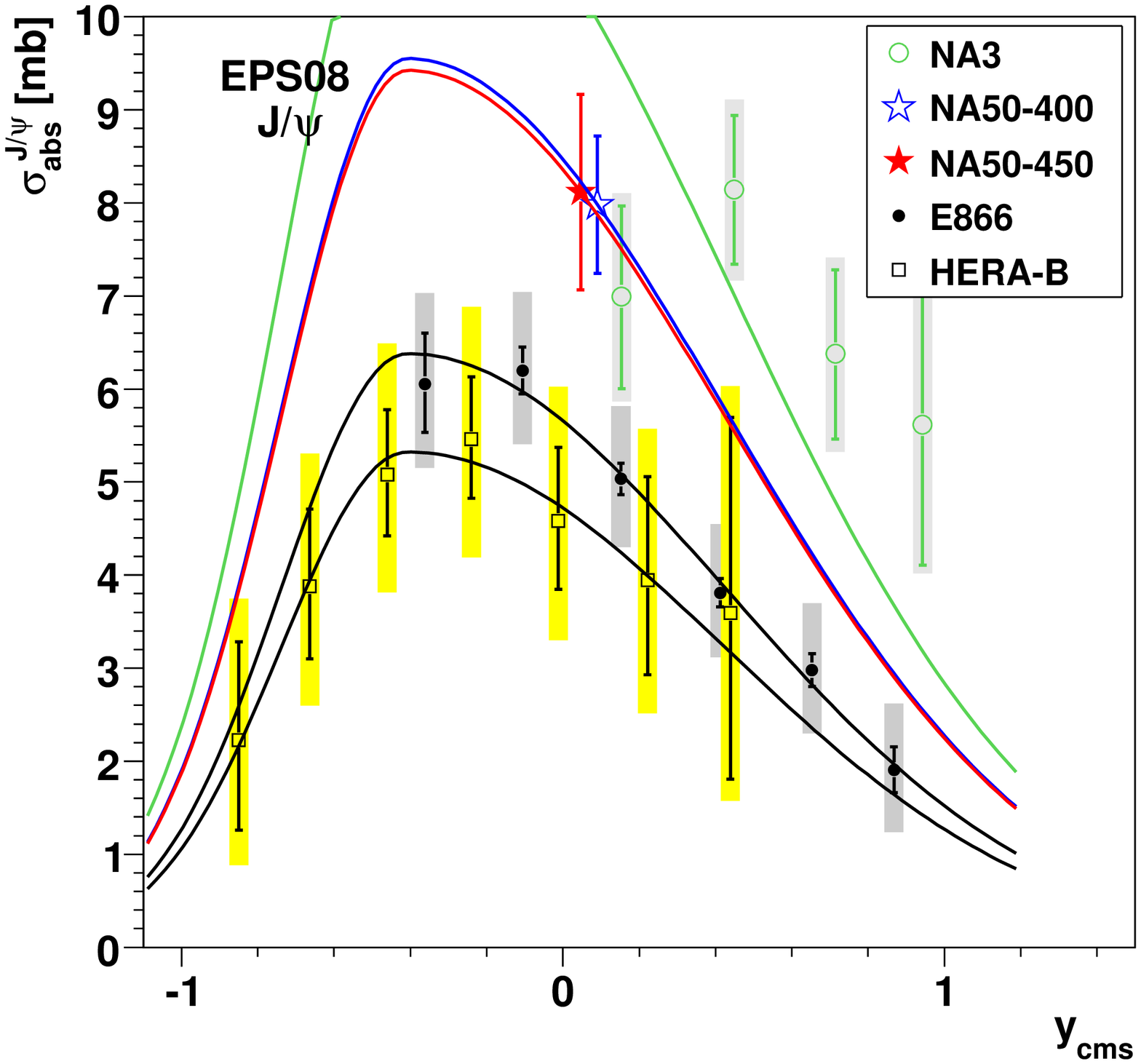}}
\caption{\sabsjpsi\ as a function of $y_{\rm cms}$, obtained by
  considering four alternative parametrisations of the nuclear PDFs.
  The boxes represent the total errors.  When nuclear PDFs are used,
  the E866 and HERA-B absorption patterns clearly depend on $y_{\rm
    cms}$.}
\label{fig:ycms}
\end{figure}

The $\sigma_{\rm abs}^{\rm J/\psi}$ values collected in
Table~\ref{tab:sigmaAbs-jPsi} are shown in Fig.~\ref{fig:ycms} as a
function of the $y_{\rm cms}$ variable, for four different parametrisations
of the nuclear effects on the PDFs.
The E866, HERA-B and NA3 values derived using free proton PDFs
(top-left panel) can be considered flat in the midrapidity range $-0.3
< y_{\rm cms} < 1.0$.  Therefore, in this case (as for nDS) we can
evaluate $\sigma_{\rm abs}^{\rm J/\psi}(y_{\rm cms}$$=$$0)$ by simply
fitting each data set to a constant.

The nuclear modifications of the PDFs considerably affect the
$\sigma_{\rm abs}^{\rm J/\psi}(y_{\rm cms})$ pattern and a non-trivial
$y_{\rm cms}$ dependence is needed to evaluate the $\sigma_{\rm abs}^{\rm
  J/\psi}(y_{\rm cms}$$=$$0)$ values for the various data sets.  
%
%
Using EKS98 PDFs,
for example, the E866 and HERA-B data sets suggest the use of an asymmetric
Gaussian function with $\mu\approx-0.21$~mb, $\sigma_{_L}\approx0.37$~mb and
$\sigma_{_R}\approx1.1$~mb.
%
%
Since the NA50 \sabsjpsi\ values correspond to broad rapidity ranges,
the asymmetric Gaussian shape is weighted by the rapidity distribution
of the \emph{measured} \jpsi\ dimuons (those which contributed to the
derived \sabs\ value), provided in Ref.~\cite{Goncalo}.  However, this
convolution has almost no impact on the final result.

\begin{figure}[p]
\centering
\resizebox{0.48\textwidth}{!}{%
\includegraphics*{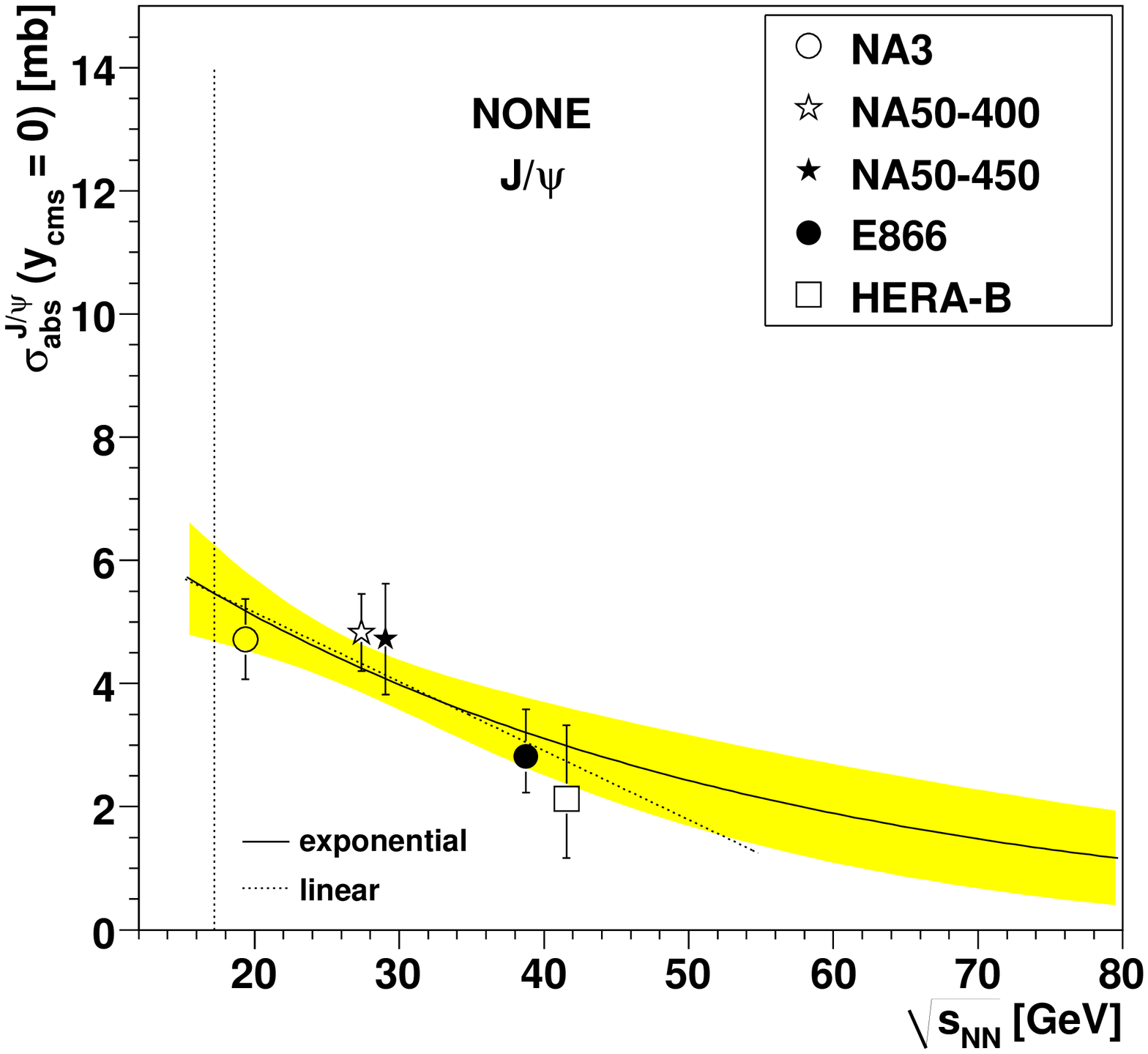}}
\resizebox{0.48\textwidth}{!}{%
\includegraphics*{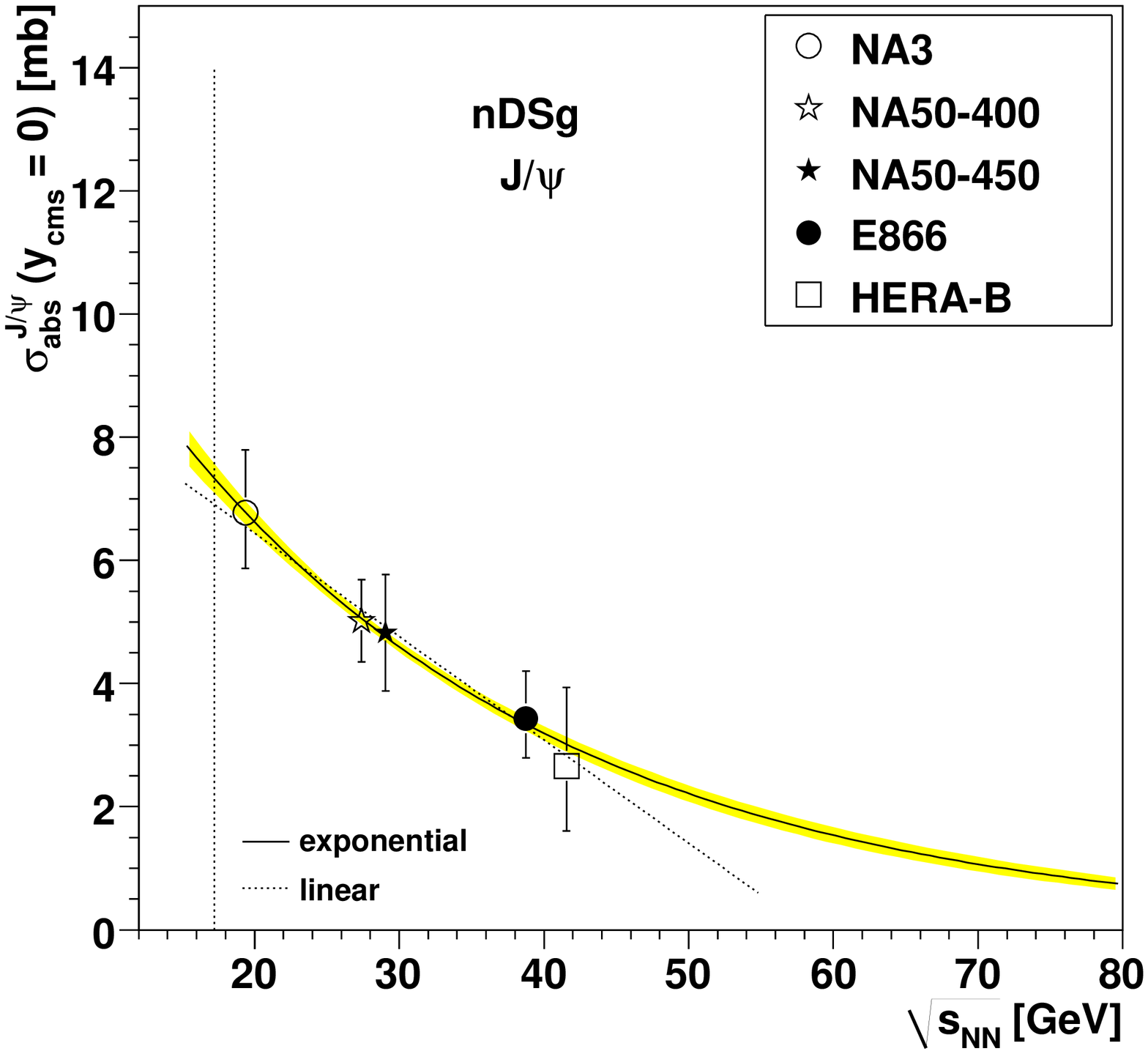}}
\resizebox{0.48\textwidth}{!}{%
\includegraphics*{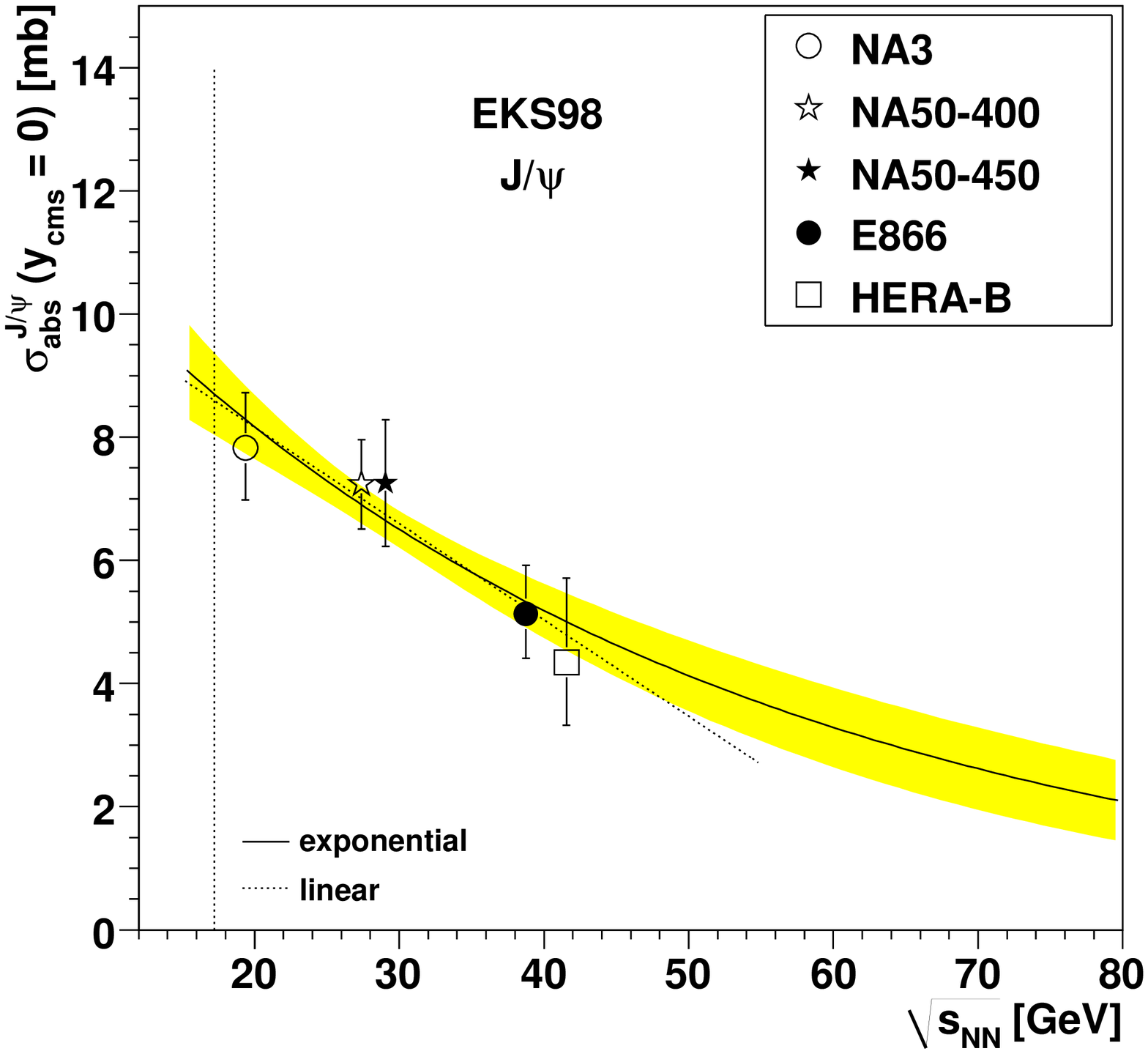}}
\resizebox{0.48\textwidth}{!}{%
\includegraphics*{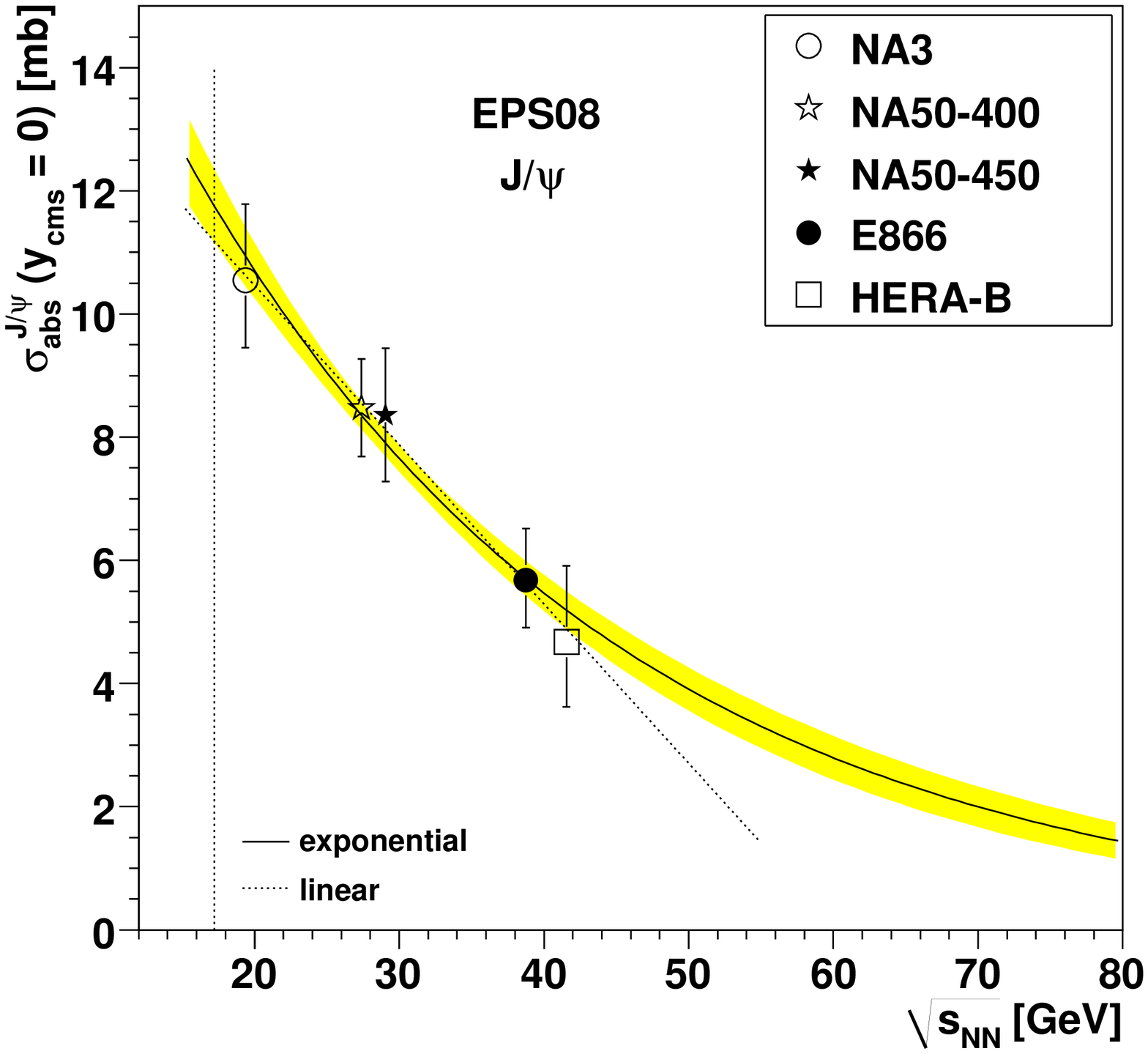}}
\caption{Dependence of $\sigma_{\rm abs}^{\rm J/\psi}(y_{\rm
    cms}$$=$$0)$ on the nucleon-nucleon centre-of-mass energy.  The
  curves represent fits with exponential (solid line with error band)
  and linear (dotted line) functions.}
\label{fig:sigma_sqrts_nn}
\end{figure}

The change in the magnitude of $\sigma_{\rm abs}^{\rm J/\psi}(y_{\rm
  cms}$$=$$0)$ with collision energy, \ssnn, can be observed in
Fig.~\ref{fig:sigma_sqrts_nn}, for free proton PDFs (``NONE'') and for
the nDSg, EKS98 and EPS08 parametrisations of the nuclear PDFs.  The
corresponding numerical values are collected in
Table~\ref{tab:sigmaycms0}.
To determine the \sabsjpsi\ relevant for the analysis of the SPS
heavy-ion results, we extrapolate $\sigma_{\rm abs}^{\rm
  J/\psi}(y_{\rm cms}$$=$$0)$ down to $\sqrt{s_{_{NN}}}=17.2$~GeV
(dotted vertical line in Fig.~\ref{fig:sigma_sqrts_nn}), using
exponential and linear functions.

\begin{table}[p]
\centering
  \caption{$\sigma_{\rm abs}^{\rm J/\psi}(y_{\rm cms}$$=$$0)$ values
    extracted from the five analysed data sets and for the nuclear
    PDFs we have considered, including the free protons case.}
\label{tab:sigmaycms0}
\vglue2mm
\begin{tabular}{lccccc} \hline
\rule{0pt}{0.5cm}
Exp. & \multicolumn{5}{c}{$\sigma_{\rm abs}^{\rm J/\psi}(y_{\rm cms}$$=$$0)$ [mb]}\rule{0pt}{0.5cm}\\
             &    NONE     &    nDS      &    nDSg     &    EKS98    &    EPS08     \\ \hline
NA3      & $4.71\pm0.66$ & $4.76\pm0.66$ & $6.78\,^{+\,1.01}_{-\,0.91}$ & $7.82\,^{+\,0.90}_{-\,0.84}$ & $10.55\,^{+\,1.24}_{-\,1.10}$\\
NA50-400 & $4.82\pm0.63$ & $4.74\pm0.62$ & $5.02\pm0.67$ & $7.24\pm0.73$ & $8.48\pm0.79$\\
NA50-450 & $4.72\pm0.90$ & $4.61\pm0.90$ & $4.82\pm0.95$ & $7.25\pm1.03$ & $8.36\pm1.08$\\
E866     & $2.82\,^{+\,0.76}_{-\,0.59}$ & $2.53\,^{+\,0.75}_{-\,0.63}$ & $3.43\,^{+\,0.77}_{-\,0.64}$ & $5.13\,^{+\,0.79}_{-\,0.72}$ & $5.68\,^{+\,0.84}_{-\,0.77}$\\
HERA-B   & $2.13\,^{+\,1.19}_{-\,0.96}$ & $1.93\,^{+\,1.15}_{-\,0.97}$ & $2.66\,^{+\,1.28}_{-\,1.05}$ & $4.35\,^{+\,1.37}_{-\,1.03}$ & $4.67\,^{+\,1.24}_{-\,1.05}$\\ \hline
\end{tabular}
\end{table}
\begin{figure}[ht!]
\centering
\resizebox{0.5\textwidth}{!}{%
\includegraphics*{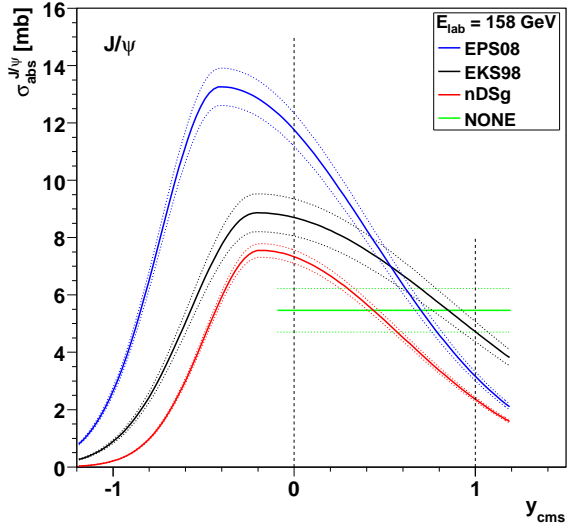}}
\caption{\sabsjpsi\ as a function of $y_{\rm cms}$, extrapolated to
  158~GeV considering four alternative parametrisations of the nuclear
  PDFs.}
\label{fig:158}
\end{figure}

\begin{table}[h!]\centering
  \caption{The $\sigma_{\rm abs}^{\rm J/\psi}(y_{\rm cms}$$=$$0)$ and
    $\sigma_{\rm abs}^{\rm J/\psi}(0<y_{\rm cms}<1)$ values evaluated
    at 158~GeV by extrapolating with an exponential function the
    values derived from measurements made at higher energies, by
    NA3, NA50, E866 and HERA-B.}
\label{tab:results}
\vglue2mm
\begin{tabular}{lcc} \hline
N-PDFs & $\sigma_{\rm abs}^{\rm J/\psi}(y_{\rm cms}$$=$$0)$ [mb]
& $\sigma_{\rm abs}^{\rm J/\psi}(0<y_{\rm cms}<1)$ [mb]\rule{0pt}{0.55cm}\\ \hline
NONE   &  $5.5\pm0.8$ & $5.5\pm0.8$ \\
nDS    &  $5.6\pm0.8$ & $5.6\pm0.8$ \\
nDSg   &  $7.3\pm0.2$ & $5.2\pm0.2$ \\
EKS98  &  $8.7\pm0.7$ & $7.2\pm0.5$ \\
EPS08  & $11.8\pm0.6$ & $7.5\pm0.4$ \\ \hline
\end{tabular}
\end{table}

The dependence of $\sigma_{\rm abs}^{\rm J/\psi}$ on $y_{\rm cms}$
at $\sqrt{s_{_{NN}}}=17.2$~GeV is shown in Fig.~\ref{fig:158},
for several nuclear PDFs, while the
values of $\sigma_{\rm abs}^{\rm J/\psi}(y_{\rm cms}$$=$$0)$
are collected in Table~\ref{tab:results}.  
%
Integrating these functions in the NA50 heavy-ion rapidity window at
158~GeV, $0 <y_{\rm cms}<1$, weighted by the rapidity distribution of
the \jpsi\ dimuons measured in Pb-Pb collisions (before acceptance
corrections)~\cite{Helena}, we obtain the values of $\sigma_{\rm abs}^{\rm
  J/\psi}(0<y_{\rm cms}<1)$ given in Table~\ref{tab:results}.

\subsection{\boldmath Dependence of \sabsjpsi\ on the \jpsi\ energy}

We now consider a different scenario.  If we assume that the \jpsi\ is
``broken up'' by interactions with nucleons while traversing the
nuclear target, it is natural to study the absorption as a function of
the \jpsi-nucleon centre-of-mass energy, \sspsi, a quantity that
reflects both the nucleon-nucleon centre-of-mass energy, \ssnn, and
the energy of the \jpsi.  This energy may be written as a function of
the \jpsi\ \xf\ as
\begin{equation}
  \label{eq:sqrts-full}
  \sqrt{s_{_{\psi\,N}}} = \sqrt{m_{\psi}^2 + \sqrt{s_{_{NN}}}~
    \Bigg( x_{\rm F}\, \sqrt{\frac{s_{_{NN}}}{4} - m_{\psi}^2} +
    \sqrt{p_{\rm T}^2 + x_{\rm F}^2\,\big( \frac{s_{_{NN}}}{4}-m_{\psi}^2\big)
      + m_{\psi}^2}\Bigg) } \quad ,
\end{equation}
where $m_{\psi}$ and $p_{\rm T}$ are the mass and transverse momentum
of the \jpsi.  When $x_{\rm F} = 0$ and the \pt\ dependence is
neglected, this expression reduces to
\begin{equation}
\sqrt{s_{_{\psi\,N}}} = m_{\psi} ~ \sqrt{1 + \frac{\sqrt{s_{_{NN}}}}{m_{\psi}}}
\quad .
\end{equation}

\begin{figure}[h!]\centering
\resizebox{0.48\textwidth}{!}{%
\includegraphics*{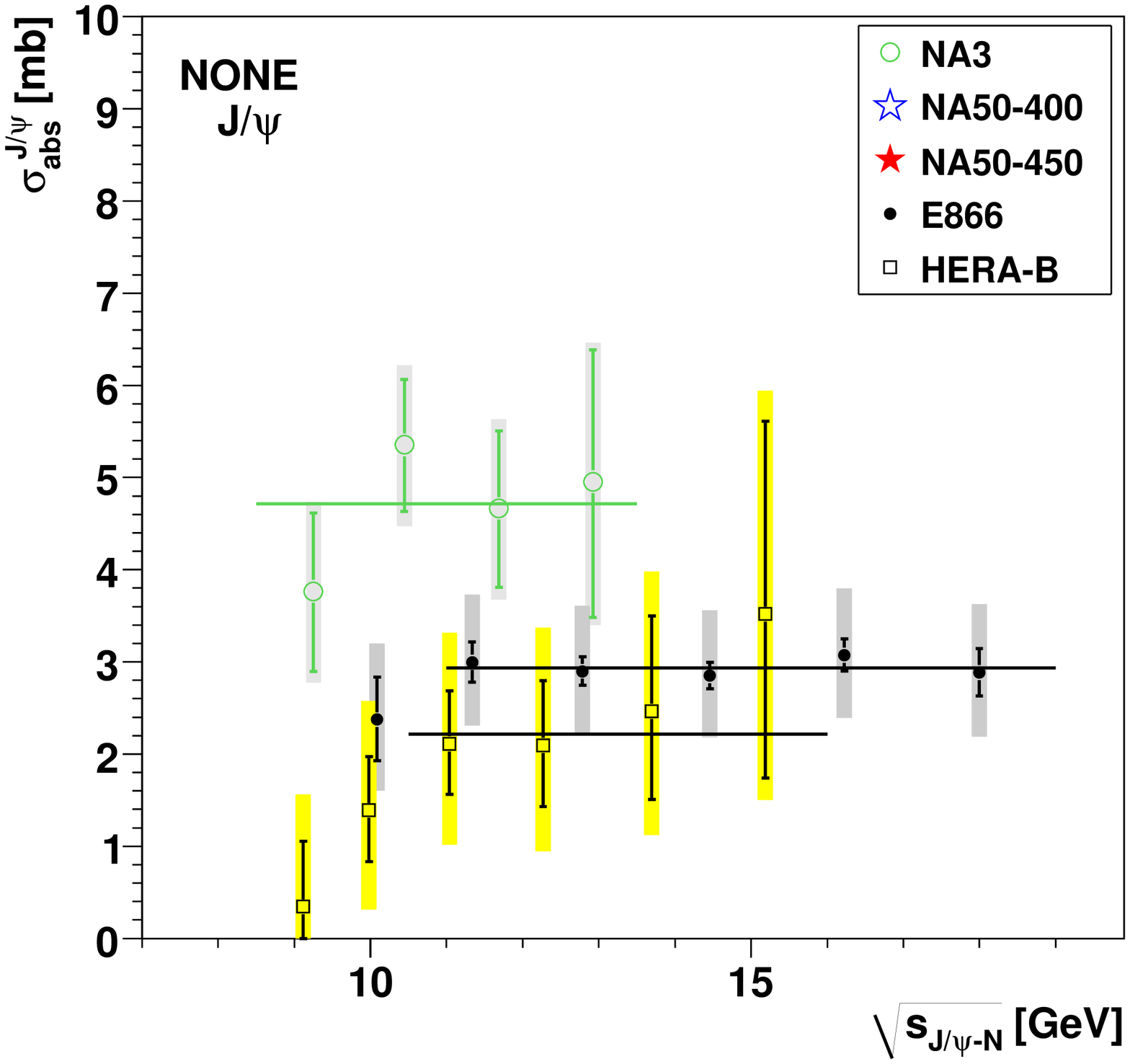}}
\resizebox{0.48\textwidth}{!}{%
\includegraphics*{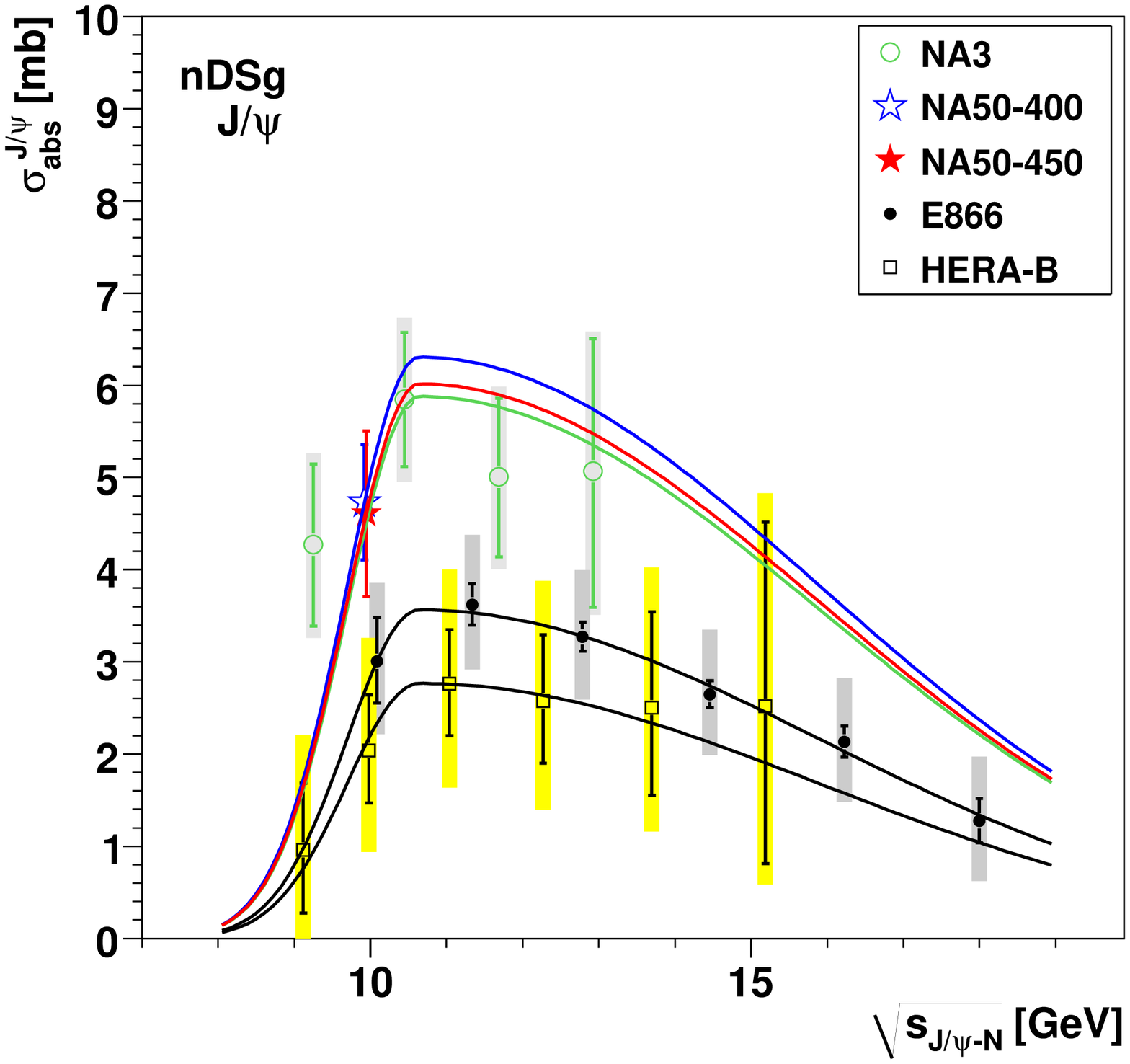}}
\resizebox{0.48\textwidth}{!}{%
\includegraphics*{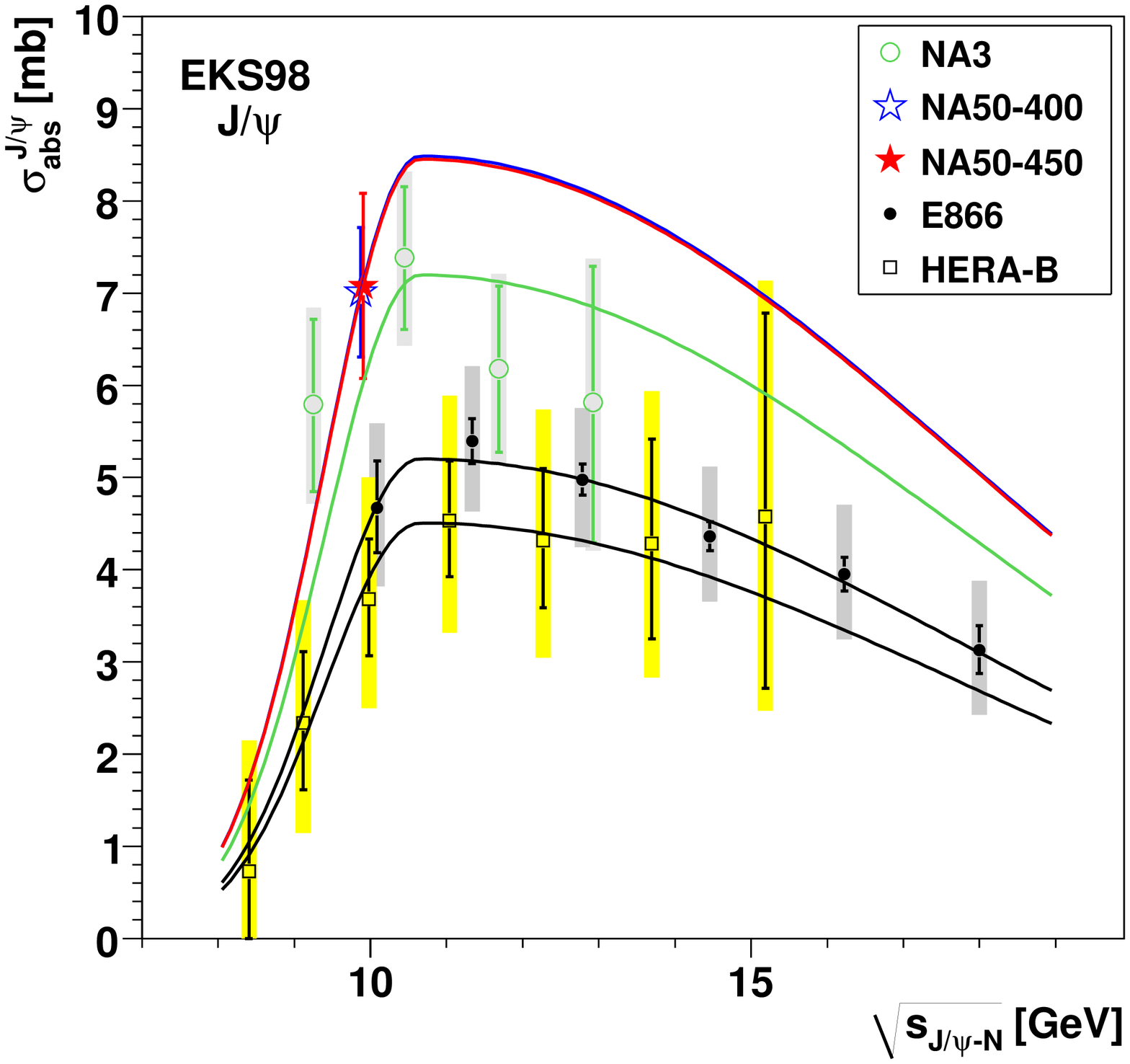}}
\resizebox{0.48\textwidth}{!}{%
\includegraphics*{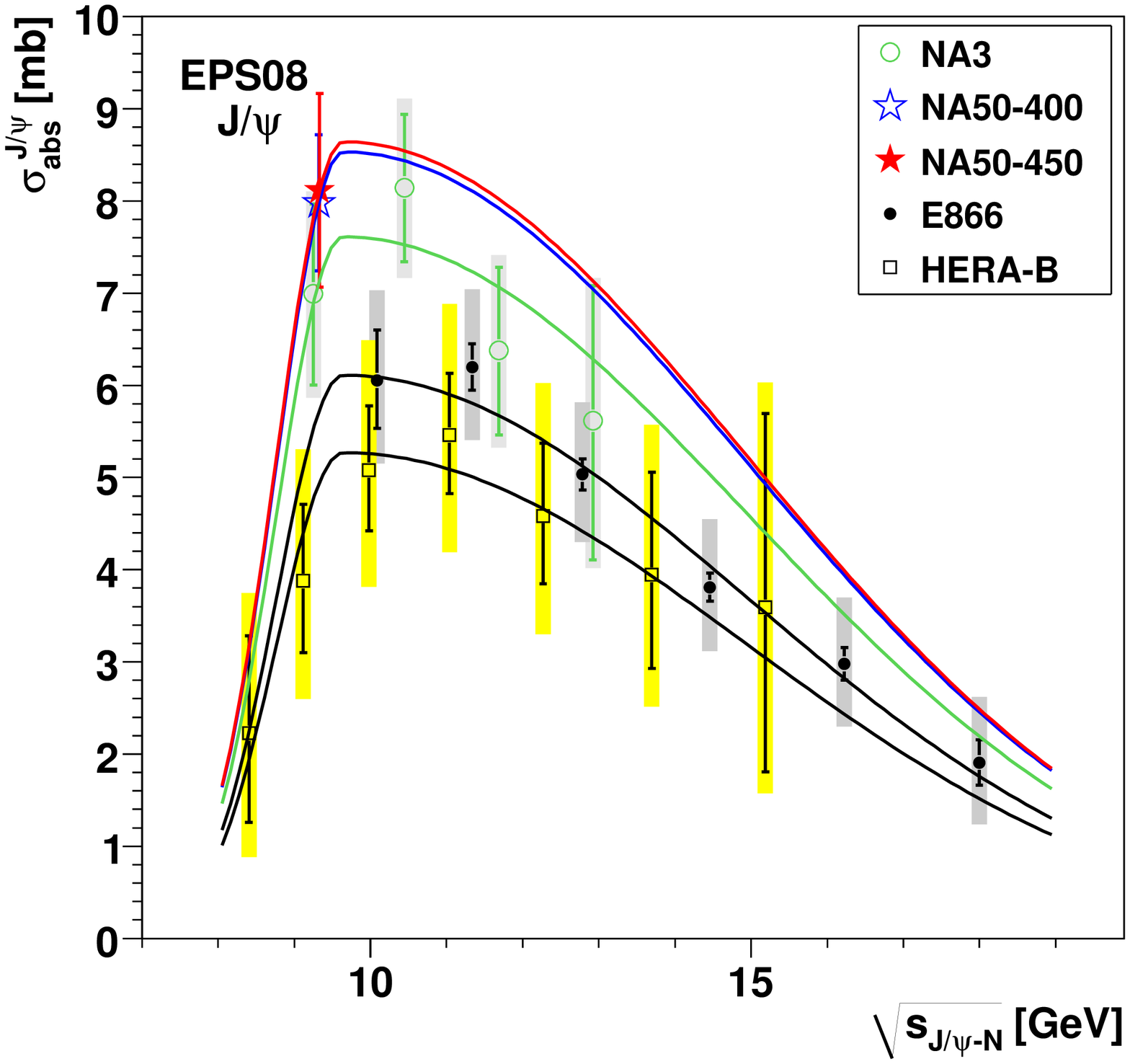}}
\caption{$\sigma_{\rm abs}^{\rm J/\psi}$ as a function of the
  \jpsi-nucleon centre-of-mass energy, determined from the
  fixed-target data sets, with four different nuclear PDFs.  The boxes
  represent the total errors.}
\label{fig:sigma_sqrts_psin}
\end{figure}

Figure~\ref{fig:sigma_sqrts_psin} shows \sabsjpsi\ as a function of
\sspsi\ for the fixed-target data sets.  The centres and widths of the
asymmetric Gaussians are exclusively determined by the E866
and \mbox{HERA-B} points while the magnitudes are independently fitted
for each data set.  The NA3 values are closer to the others here than
as a function of $y_{\rm cms}$ (Fig.~\ref{fig:ycms}).  However, it is clear that
there is also no scaling as a function of \sspsi\  since the different
data sets result in significantly different magnitudes of \sabsjpsi;
the E866 and \mbox{HERA-B} values are too low while NA50 is too high relative
to a global average.
The decrease of \sabsjpsi\ with increasing beam energy is always seen,
regardless of the nuclear PDF set, including that of the free proton (NONE).

The absence of a universal curve indicates that the mechanisms
determining the \jpsi\ nuclear dependence are not accurately described
by the simple ``Glauber absorption model'' we used in our analysis.
An improved model is needed to properly explain charmonium absorption
and its dependencies on collision energy and kinematics, including the
extra absorption seen at forward \xf.

\section{Summary and outlook}

Previous derivations of the \jpsi\ normal nuclear absorption baseline
used to search for QGP signals in the SPS heavy-ion measurements,
collected at $E_{\rm lab}=158$~GeV, were based on proton-nucleus data
collected at 400--450~GeV, assuming that \sabsjpsi\ is a ``universal
quantity'', insensitive to changes related to the collision energy or
to the rapidity window.
This paper presents an improved analysis of charmonium production in
proton-nucleus collisions.  First, we studied \jpsi\ proton-nucleus
data collected in several fixed-target experiments, covering a broad
range of collision energies ($\sqrt{s_{_{NN}}} = 20$--40~GeV), as well
as d-Au data collected by PHENIX at $\sqrt{s_{_{NN}}} = 200$~GeV
(presently affected by large uncertainties).  Second, we considered
nuclear modifications of the PDFs, employing several models which
consistently indicate initial-state gluon enhancement (antishadowing)
in the midrapidity region of the fixed-target data.

We observe that, when the nuclear modifications of the PDFs are taken
into account, \sabsjpsi\ significantly depends on the rapidity of the
\jpsi, even within a relatively narrow midrapidity window.  In
particular, the \jpsi\ nuclear dependence determined by E866 in the
window $-0.1 < x_{\rm F} < +0.2$ only looks independent of \xf\ if the
nuclear effects on the PDFs are neglected.
Furthermore, the level of cold nuclear matter absorption of
midrapidity \jpsi\ significantly decreases with collision energy.
While the specific numerical values depend on the nuclear PDF sets
used, the decrease of \sabsjpsi\ with energy is a general feature,
independent of any nuclear modifications of the
PDFs.
The observation that \sabsjpsi\ depends on the rapidity of the \jpsi\
and on the collision energy confirms that the simple Glauber-type
absorption model commonly used in \jpsi\ suppression studies is
insufficient to properly reproduce the available measurements.  The
sought-for ``universal quantity'' is, after all, a multidimensional
function without obvious scaling features.

For the moment, in the absence of a more detailed formalism, we used
the data-driven $\sigma_{\rm abs}^{\rm J/\psi}(y_{\rm cms})$ dependence
and a simple extrapolation of the \ssnn\ dependence to
evaluate the \sabsjpsi\ value corresponding to the rapidity window
covered by NA50 at 158~GeV, obtaining the values summarised in
Table~\ref{tab:results}.
The \sabsjpsi\ values obtained in this study should be seen as
effective ones, incorporating the nuclear matter effects on 
the directly produced \jpsi\ as well as on the more massive
$\chi_{c}$ and \psip\ states.  Indeed, around 33\,\% of the
observed \jpsi\ yields result from decays of $\chi_{c}$ and \psip\
states~\cite{feeddown}, which are expected to suffer stronger nuclear
absorption than the directly produced \jpsi.

Having relaxed the assumption of a universal absorption cross section,
we can test whether it is justified to study the SPS heavy-ion data,
collected at 158~GeV, using the $\sigma_{\rm abs}^{\rm J/\psi}$ value
obtained from the higher-energy NA50 measurements.
As shown in Table~\ref{tab:results}, if we neglect nuclear
modifications on the PDFs we derive $\sigma_{\rm abs}^{\rm
  J/\psi}(0<y_{\rm cms}<1)= 5.5\pm0.8$~mb, higher than the value used
so far in the analyses of the SPS heavy-ion measurements
($4.2\pm0.5$~mb, also using free proton PDFs).  If we use the EKS98
parametrisation to model the nuclear modifications of the PDFs, we
obtain $\sigma_{\rm abs}^{\rm J/\psi}(0<y_{\rm cms}<1)= 7.2\pm0.5$~mb.
It is interesting to notice that this is identical to the value
obtained by NA50 at 400~GeV in the rapidity window $-0.425 < y_{\rm
  cms} < 0.575$ ($7.01\pm0.70$~mb, see Table~\ref{tab:sigmaAbs-jPsi}).
The drop of \sabsjpsi\ from $y_{\rm cms} = 0$ to $y_{\rm cms} = 1$, in
the EKS98 case, exactly compensates the increase in \sabsjpsi\ from
$E_{\rm lab}=400$ to 158~GeV.
Nevertheless, a quantitative re-evaluation of the level of ``QGP
melting'' in the heavy-ion data should be performed, also considering
nuclear modifications of the PDFs in the beam nucleus (in an $x$ range
different from that of the target nucleus).

Our understanding of the cold nuclear effects on \jpsi\ production,
and their energy dependence, should significantly improve in the near
future, thanks to new NA60 results based on proton-nucleus data
collected at 158~GeV and to new PHENIX results based on a large d-Au
data set.

\section*{Acknowledgements}

We acknowledge fruitful discussions with Gon\c{c}alo Borges and Helena
Santos (NA50), Mike Leitch (E866), Philippe Charpentier (NA3), and
Pietro Faccioli (\mbox{HERA-B}).
Carlos Salgado and Kari Eskola provided insightful information on
parton densities in heavy nuclei.
This study reflects many interesting discussions, over many years,
with Roberta Arnaldi, Dima Kharzeev, Louis Kluberg, Helmut Satz,
Enrico Scomparin, and Jo\~ao Seixas.
Finally, it is a pleasure to thank the JHEP referee, whose feedback
contributed to improve the clarity and robustness of our paper.

The work of R.V.\ was performed under the auspices of the U.S.\
Department of Energy by Lawrence Livermore National Laboratory under
Contract DE-AC52-07NA27344 and was also supported in part by the
National Science Foundation Grant NSF PHY-0555660.  The work of
H.K.W.\ was supported by the Portuguese Funda\c{c}\~ao para a
Ci\^encia e a Tecnologia, under Contract SFRH/BPD/42138/2007.


\end{document}